\documentclass[twoside,leqno]{article}

\usepackage[letterpaper]{geometry}

\usepackage{siamproceedings}

\usepackage[T1]{fontenc}
\usepackage{amsfonts,tikz}
\usepackage{graphicx,xcolor}
\usepackage{epstopdf,subfig}
\usepackage{enumitem}
\usepackage{algorithmic}
\ifpdf
  \DeclareGraphicsExtensions{.eps,.pdf,.png,.jpg}
\else
  \DeclareGraphicsExtensions{.eps}
\fi

\newsiamremark{remark}{Remark}
\newsiamremark{hypothesis}{Hypothesis}
\crefname{hypothesis}{Hypothesis}{Hypotheses}
\newsiamthm{claim}{Claim}

\usepackage{amsopn}

\newcommand{\C}{\mathbb{C}}

\newcommand{\D}{\mathbb{D}}
\newcommand{\R}{\mathbb{R}}
\newcommand{\Z}{\mathbb{Z}}
\newcommand{\N}{\mathbb{N}}

\newcommand{\A}{\mathbb{A}}
\newcommand{\dd}{\text{\rm d}}        
\newcommand{\E}{\mathbb{E}}

\newcommand{\mc}{\mathcal}
\def\T{\mathbb{T}}
\newcommand{\bs}{\boldsymbol}
\begin{document}

\newcommand\relatedversion{}
\renewcommand\relatedversion{\thanks{The full version of the paper can be accessed at \protect\url{https://arxiv.org/abs/0000.00000}}} 

\title{\Large Two Decades of Probabilistic Approach to Liouville Conformal Field Theory\relatedversion}
    \author{R\'emi Rhodes\thanks{Aix-Marseille University, CNRS, I2M, Marseille, France, \email{remi.rhodes@univ-amu.fr}}
    \and Vincent Vargas\thanks{Section de math\'ematiques et d\'epartement  de physique th\'eorique, University of Geneva, Geneva,
  \email{Vincent.Vargas@unige.ch}.}}

\date{}

\maketitle


\fancyfoot[R]{\scriptsize{Copyright \textcopyright\ 20XX by SIAM\\
Unauthorized reproduction of this article is prohibited}}





\begin{abstract} Over the past twenty years, the probabilistic approach to Liouville Conformal Field Theory (LCFT) has undergone remarkable developments, transforming a collection of ideas at the interface of probability, geometry, complex analysis and physics into a coherent mathematical theory. Building on Gaussian Free Fields and Gaussian Multiplicative Chaos, rigorous definitions of correlation functions and partition functions have been established, culminating in the probabilistic derivation of the DOZZ formula and a mathematically complete formulation of the conformal bootstrap  for LCFT on Riemann surfaces. This survey aims to provide a synthetic account of these advances, emphasizing both the main achievements and the open problems that continue to drive the field.
\end{abstract}

 \section{Introduction.}
 
Liouville Conformal Field Theory (LCFT) occupies a central position in two-dimensional quantum field theory and statistical physics. Originally introduced by Polyakov in 1981 \cite{Polyakov81} as a model for fluctuating two-dimensional geometries, Liouville theory plays a key role in string theory, random geometry, and the theory of two-dimensional quantum gravity. 

From the physics perspective, given a two dimensional Riemannian manifold $(\Sigma,g)$,  LCFT is defined via a path integral  (namely a formal integration measure over a functional space)
\begin{equation}\label{intro:path}
\int e^{-S_g(\phi)} \, \mathcal{D}\phi,
\end{equation}
with
\begin{equation}\label{intro:action}
S_g(\phi):=\frac{1}{4\pi}\int_\Sigma \big(|d \phi|^2_g + Q K_g \phi + 4\pi \mu e^{\gamma \phi}\big)\, \dd\mathrm{v}_g
\end{equation}
where the formal integration measure $\mathcal{D}\phi$ stands for the putative Lebesgue measure on the space of maps $\phi:\Sigma\to\mathbb{R}$, $\gamma \in (0,2)$ is the coupling constant, $Q = \frac{2}{\gamma} + \frac{\gamma}{2}$ is the background charge, $K_g$ is the scalar curvature, ${\rm v}_g$ the volume form and $\mu>0$ is called the cosmological constant. 

The challenge of giving rigorous meaning to this formal path integral has motivated probabilists and analysts for decades. Over the last twenty years, remarkable progress has been achieved, resulting in a fully rigorous construction of LCFT correlation functions,   the derivation of the celebrated DOZZ formula \cite{DornOtto94,Zamolodchikov96} and of the conformal bootstrap \cite{BPZ84,Zamolodchikov96}. These advances are based on probabilistic tools: Gaussian Free Fields (see \cite{Sheffield07} for a review), Gaussian Multiplicative Chaos \cite{Kahane85}, and stochastic processes describing random geometry, combined with methods from complex analysis, scattering theory and geometry of moduli spaces.

The aim of this article is to provide a panoramic account of these developments, accessible to a broad mathematical audience. We focus primarily on the probabilistic techniques and their implications for both probability theory and mathematical physics. For further details, the reader may consult the comprehensive review \cite{GKRreview}.

 \section{A brief history of Liouville CFT.}

In his seminal 1981 paper \emph{Quantum geometry of bosonic strings} \cite{Polyakov81}, Polyakov initiated the study of two-dimensional quantum gravity by proposing to sum over random metrics on a given Riemann surface $\Sigma$, thus describing a random metric on this surface (see also \cite{David88,Distler_Kawai}). His crucial observation was that, conditionally on the moduli parameter, the law of the conformal factor of the metric should be governed by the Liouville path integral \cref{intro:path}, where $g$ is any arbitrarily fixed representative of the conformal class. This path integral describes a random metric $e^{\gamma\phi}g$ fluctuating around the critical points of the classical Liouville action, namely metrics with uniformized curvature on~$\Sigma$.

The challenge was then to compute expectation values of the natural observables of the theory, namely the correlation functions of vertex operators (see \cref{actioninsertion} below). Probabilistically, these correlations play the role of a Laplace transform characterizing the path integral, while geometrically they force the random metric to fluctuate around metrics with uniformized curvature outside a given set of points, with further conical singularities at those points.  

Since a direct evaluation of these correlation functions proved to be intractable, Belavin, Polyakov,  and Zamolodchikov introduced in 1984 the \emph{conformal bootstrap} \cite{BPZ84}, a new paradigm to solve two-dimensional conformal field theories (CFTs). The bootstrap is based on conformal symmetry and the operator product expansion (OPE), which impose recursive algebraic constraints on correlation functions. This method led to spectacular progress in the classification of rational CFTs, but remained inapplicable to Liouville theory, as the recursion requires the three-point function, then unknown. This prompted Polyakov to call CFT an ``\textit{unsuccessful attempt to solve the Liouville model}'', and he initially hesitated to publish the work, see \cite{polyakovquark}. While powerful, the bootstrap is essentially an algebraic framework, and does not directly connect to the path integral picture or to statistical mechanics models at criticality (such as the critical Ising model), which are expected to converge to CFTs in the scaling limit.

A decade later, a formula for the  missing three-point function of Liouville theory was proposed independently by Dorn-Otto and Zamolodchikov-Zamolodchikov \cite{DornOtto94,Zamolodchikov96}, based on ingenious manipulations of the Liouville path integral and Coulomb-gas techniques. Their formula, now known as the DOZZ structure constant, provided a consistent solution to the bootstrap equations. The derivation was later significantly refined by Teschner \cite{Tesc}, using BPZ differential equations and analytic continuation. Yet the relation between the bootstrap and the path integral (still lacking a rigorous definition) remained mysterious.

Meanwhile, two mathematical frameworks were developed to formalize the bootstrap. On the algebraic side, \emph{vertex operator algebras} (VOAs) \cite{Borcherds,Frenkel:1988xz,Huang1997TwoDimensionalCG} provided an axiomatic setting for the operator product expansion. On the geometric side, Segal proposed in 1987 a set of axioms \cite{Segal87} inspired by the path integral picture, where correlation functions are obtained by gluing Riemann surfaces equipped with boundary Hilbert spaces. Yet, for decades, Liouville theory remained beyond reach of either formalism.

Recent works have closed this gap. Using probabilistic tools --Gaussian free fields \cite{Sheffield07}, Gaussian multiplicative chaos \cite{Kahane85}, and stochastic analysis-- the Liouville path integral has now been constructed rigorously on any Riemann surface \cite{DKRV16,DRV16_tori,huang2018,GRV2019}. This approach shows that the probabilistic path integral for Liouville theory satisfies Segal's axioms \cite{GKRV21_Segal}, which are designed to connect to the (more standard) Hilbert space picture and operator formalism in Quantum Field Theory, yielding a representation of the symmetry algebra of the Liouville theory, namely the Virasoro algebra, acting on an explicit Hilbert space. Scattering theory methods were then used in \cite{GKRV20_bootstrap} to diagonalize the generator of dilations (the Hamiltonian) and decompose the Hilbert space as a sum of irreducible Virasoro representations. This led to a proof of the conformal bootstrap \cite{GKRV20_bootstrap,GKRV21_Segal}, with the DOZZ formula rigorously established in \cite{KRV19_local,KRV_DOZZ}. This work provides an instance of a non-trivial CFT where the path integral, Segal's axioms, operator formalism and conformal bootstrap are realized consistently within a complete mathematical framework.

\begin{figure} 
\centering
\subfloat[Random triangulation of the sphere]{\includegraphics[width=8cm]{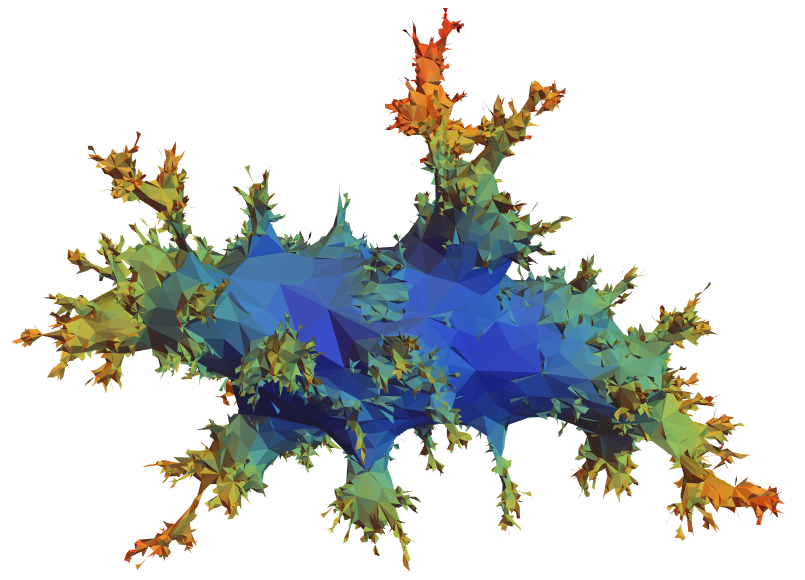}}%
\qquad
\subfloat[Local zoom in the triangulation once conformally embedded into the Riemann sphere]{\includegraphics[width=6.8cm,height=5.7cm]{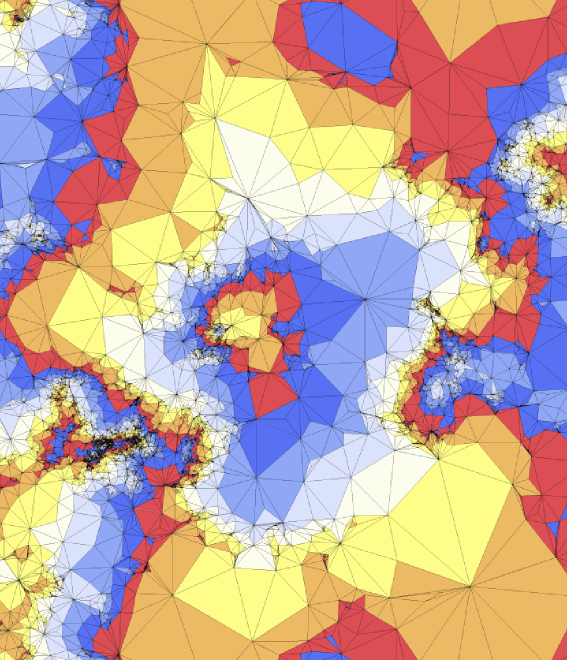}}
\caption{Triangulation of the Riemann sphere sampled from the uniform probability measure.@T.Budd}%
\label{triang}%
\end{figure}

Beyond its role as a canonical conformal field theory, Liouville theory also emerges naturally in probability and statistical physics. In particular, in the context of two-dimensional quantum gravity, Liouville theory is conjecturally equivalent to the scaling limit of random planar maps, i.e.\ probability measures on finite triangulations of a fixed topological surface, see \Cref{triang}. This conjecture has been proven in the case of uniform planar maps through the works \cite{LeGall14,Miller-Sheffield1,Miller-Sheffield2,Miller-Sheffield3,holdencardy}, while the general equivalence remains an active direction of research (see also \cite[Appendix 5.3]{DKRV16}).

\section{Probabilistic foundations.}\label{sec:proba}

Classically, when $Q$ takes the value $Q=\tfrac{2}{\gamma}$, the critical points of the action functional \eqref{intro:action} correspond to metrics conformally equivalent to the reference metric $g$ with uniform curvature. Concretely, they are of the form, for $u\in C^\infty(\Sigma)$,
\begin{equation}
g'=e^{\gamma u}g,\qquad K_{g'}=-2\pi\mu\gamma^2.
\end{equation}
If such a metric exists, it is unique. Hence the path integral \eqref{intro:path} should be thought of as describing fluctuations of metrics around these critical points. This correspondence has been analyzed through large deviation techniques in \cite{LACOIN2017875,AFST_2022_6_31_4_1031_0} for negatively curved surfaces. The case of positive curvature remains open.

The action \eqref{intro:action} is nonlinear in the field $\phi$, which complicates the construction of the path integral \eqref{intro:path}. The key first step is to reinterpret the quadratic 
part: the squared gradient can be expressed in terms of a probabilistic object called the Gaussian Free Field (GFF). This is a centered Gaussian process with covariance given by the Green function of the Laplace-Beltrami operator (see \cite{Berestycki_Powell_2025,DKRV16,Dubedat_SleFreeField}).

On a compact Riemann surface $(\Sigma,g)$, the GFF can be constructed as a random series. Choose an orthonormal basis $(e_j)_{j\geq 0}$ of eigenfunctions of $\Delta_g$ with eigenvalues $(\lambda_j)_{j\geq 0}$ (with $\lambda_0=0$). Let $(a_j)_j$ be i.i.d. standard Gaussian random variables,  defined on some probability space $(\Omega,\mc{F},\mathbb{P})$. The GFF on $(\Sigma,g)$ is then the random series
\begin{equation}\label{GFFclosed}
X_g(x) := \sqrt{2\pi}\sum_{j\ge 1} \frac{a_j e_j(x)}{\sqrt{\lambda_j}} .
\end{equation}
This series converges in the Sobolev space   $H^s(\Sigma):=(1+\Delta_g)^{-s/2}(L^2(\Sigma))$ for any $s<0$. Therefore, it is not a fairly defined function, but rather a distribution in the sense of Schwartz. The GFF has expectation zero, and covariance $\E[X_g(x)X_g(y)]=2\pi G_g(x,y)$ where $G_g$ is the Green function of the Laplacian with zero ${\rm v}_g$-mean condition. Thus the quadratic part of the action defines a Gaussian measure:
\begin{equation}\label{mesureGFFint}
\int F(\phi)e^{-\frac{1}{4\pi}\int_\Sigma|d\phi|_g^2 {\rm dv}_g} D\phi :=
\Big( \frac{{\rm v}_g(\Sigma)}{{\det}'(\Delta_g)} \Big)^{1/2}\int_\R \E[F(c+X_g)] \dd c.
\end{equation}
Here ${\det}'(\Delta_g)$ is the regularized determinant of the Laplacian \`a la Ray-Singer \cite{Ray-Singer}, and the integral over $c$ accounts for the constant mode (absent from \eqref{GFFclosed}).

\begin{figure} 
\centering
\includegraphics[width=8cm]{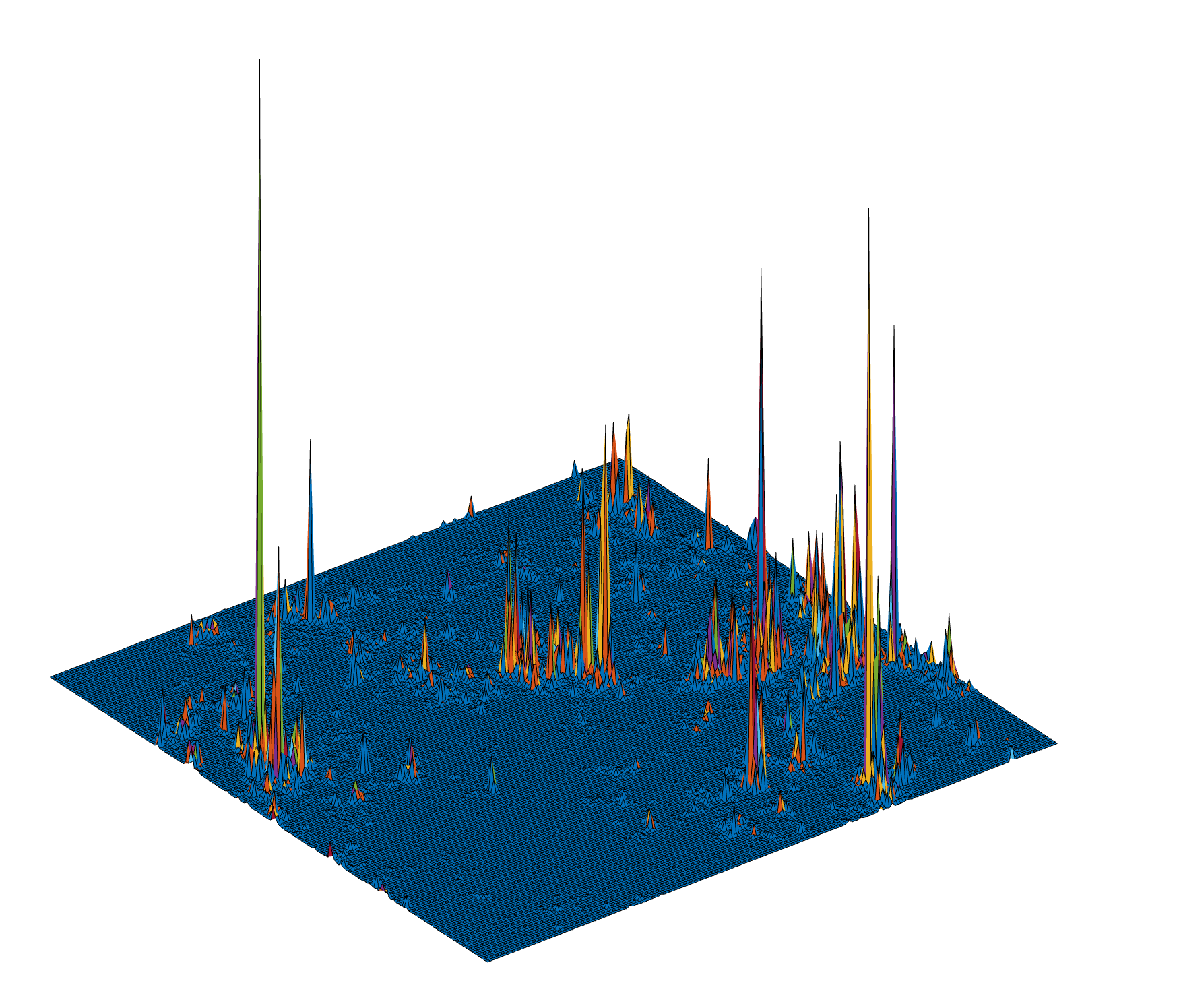}
\caption{A GMC measure sampled on the square with $\gamma=1.1$}%
\label{GMC11}%
\end{figure}

The remaining part of the Liouville action appears as a Radon-Nikodym derivative with respect to this Gaussian measure. The curvature term is linear and harmless. The obstacle is the exponential of the GFF $\int_\Sigma e^{\gamma \phi} \dd {\rm v}_g$, since the GFF is a distribution, not a function. The way around it is Gaussian Multiplicative Chaos (GMC) theory, initiated by Kahane \cite{Kahane85} in the eighties, with roots in Mandelbrot's multiplicative cascades, and since then greatly developed (see e.g.\ \cite{Berestycki17,DuplantierSheffield11,rhodes2014_gmcReview,SHAMOV20163224}). The idea is to regularize  the GFF by convolution at scale $\epsilon$, yielding smooth fields $(X_{g,\epsilon})_{\epsilon>0}$ with covariance
\begin{equation}
G_{g,\epsilon}(x,y)=\ln\frac{1}{d_g(x,y)+\epsilon}+O(1)
\end{equation}
 with $d_g$ the distance in the metric $g$. The associated random measure on $\Sigma$  is then defined by  
\begin{equation}\label{GMC}
M^g_{\gamma}(\dd x):=\lim_{\epsilon\to 0}\epsilon^{\gamma^2/2}e^{\gamma X_{g,\epsilon}(x)}\,  {\rm v}_g(\dd x),
\end{equation}
the limit holding in probability  for the topology of weak convergence of measures on $\Sigma$. This measure is nontrivial if and only if $\gamma\in(0,2)$, which explains the constraint on $\gamma$. Moreover, its law is universal: it does not depend on the particular regularization chosen. Intuitively, $M^g_\gamma$ can be seen as the random volume form associated with the ``formal'' metric $e^{\gamma X_g}g$, see \Cref{GMC11}. The reader may consult \cite{Berestycki_Powell_2025,GRV2019} for a pedagogical introduction to the construction of these random measures.

The theory exhibits a phase transition at $\gamma=2$, where a modified renormalization still yields a nontrivial limit \cite{DRSV14_mart,DRSV14,LacoinCrit}, and universality has been studied in \cite{JunnilaSkasman17_uniqueness} (see also the review \cite{Powell20_review}). The measure $M^g_\gamma$ is the prototype of multifractal random geometry: its associated Brownian motion has been introduced in \cite{Garban16LBM,BerestLBM}, but the precise structure of its heat kernel is still unresolved despite progress \cite{DingZeitouniZhang19}. The construction of the associated random distance was achieved in \cite{DingDDF19_tightness,GwynneMiller19_metric}, with further developments in \cite{DubedatFGPS19_metric}. The central open problem related to the distance  is to determine the Hausdorff dimension of $(\Sigma, e^{\gamma X_g} g)$, known to equal $4$ when $\gamma=\sqrt{8/3}$.

 The Liouville path integral \eqref{intro:path} on $(\Sigma,g)$ can now be defined rigorously as
\begin{align}\label{defLQFT}
\langle F\rangle_{\Sigma,g} := & \Big( \frac{{\det}'(\Delta_{g})}{{\rm v}_{g}(\Sigma)} \Big)^{-1/2}  
  \times \int_\R \E\Big[ F( c+ X_{g}) \exp\Big( -\frac{Q}{4\pi}\int_{\Sigma}K_{g}(c+ X_{g} ){\rm dv}_{g} - \mu e^{\gamma c}M^g_{\gamma}(\Sigma) \Big) \Big]\dd c . \nonumber
\end{align}
for measurable $F\ge 0$ on $H^{-s}(\Sigma)$.
 
 Its total mass (the ``partition function'', obtained with $F=1$) may diverge. This can be understood via the zero mode approximation (the minisuperspace): neglecting the contribition of the GFF, one obtains (up to an overall metric-dependent factor)
$$\int_\R e^{-Q\chi (\Sigma)c-\mu e^{\gamma c}}\,\dd c$$
by Gauss-Bonnet, namely $\int_\Sigma K_g\,\dd {\rm v}_g=4\pi \chi(\Sigma)$ where $\chi(\Sigma)$ is the Euler characteristic.  This integral diverges on the sphere ($\chi(\Sigma)=2$) and torus ($\chi(\Sigma)=0$), but converges on hyperbolic surfaces ($\chi(\Sigma)\leq -2$).

Correlation functions in Liouville CFT involve the Laplace exponents $V_\alpha(x):=e^{\alpha \phi (x)}$ of the field, for $\alpha \in \R$ and $x\in \Sigma$, also called vertex operators.  Since $\phi$ is not defined pointwise, one introduces regularizations:
\begin{equation}
V_{\alpha,\epsilon}(x)=\epsilon^{\alpha^2/2}  e^{\alpha  \phi_\epsilon(x) } .
\end{equation}
For $m$ marked points ${\bf x}=(x_1,\dots,x_m)$ with weights ${\boldsymbol \alpha}=(\alpha_1,\dots,\alpha_m)$, the $m$-point correlation function is defined by
\begin{align}\label{actioninsertion}
 \langle V_{\alpha_1}(x_1)\dots V_{\alpha_m}(x_m)\rangle_{\Sigma,g}:=\lim_{\epsilon\to 0}\, \langle V_{\alpha_1,\epsilon}(x_1)\dots V_{\alpha_m,\epsilon}( x_m)\rangle_{\Sigma,g}.
\end{align}
 The limit exists exactly under Seiberg's bounds:
 
 \begin{proposition}\label{limitcorel}  If
 \begin{align}\label{seiberg1}
 & \sum_{j=1}^m\alpha_j >\chi(\Sigma)Q,\qquad \text{and }\qquad\forall j=1,\dots,m,\quad \alpha_j<Q ,
 \end{align}
then \eqref{actioninsertion} exists and is positive. Otherwise it vanishes or diverges.
\end{proposition}

On the sphere, at least three insertions are needed; on the torus, at least one. The partition function ($m=0$) is only defined on hyperbolic surfaces. Geometrically, vertex operators may be viewed as quantum analogs of conical singularities. The Seiberg bounds reflect classical conditions for existence of negatively curved metrics with prescribed conical singularities \cite{Troyanov}. The first condition   can be thought of as a probabilistic Gauss-Bonnet
theorem and can be seen in the minisuperspace as before. Probabilistically, the second bound $\alpha<Q$ ensures that the singularity induced by $V_\alpha$ is integrable with respect to $M_\gamma^g$. The multifractal analysis of $M_\gamma^g$ determines this condition (see \cite{DKRV16,rhodes2014_gmcReview}).


Correlation functions also satisfy conformal symmetry properties:
 \begin{proposition} \label{diffweyl}
\label{covconf2} Assume the Seiberg bounds are satisfied.  
\begin{enumerate}
\item {\bf Weyl anomaly:} For  each $\omega\in C^\infty(\Sigma)$,
\[ \frac{\langle V_{\alpha_1}(x_1)\dots,V_{\alpha_m}(x_m)\rangle_{\Sigma,e^\omega g}}{ \langle V_{\alpha_1}(x_1)\dots,V_{\alpha_m}(x_m)\rangle_{\Sigma,g} }=\exp\Big(\frac{c_{\rm L}}{96\pi}\int_{\Sigma}(|d\omega|_g^2+2K_g\omega) {\rm dv}_g -\sum_{j=1}^{m}\Delta_{\alpha_j}\omega(x_j)\Big)\]
where $c_{\rm L}=1+6Q^2$ is called the central charge  and the real numbers $\Delta_{\alpha_i}$, called {\it conformal weights}, are defined by the relation $\Delta_{\alpha}:=\frac{ \alpha}{2}(Q-\frac{\alpha}{2}) $ for $\alpha\in\R$.
\item {\bf diffeomorphism invariance:} Let   $\psi:\Sigma'\to \Sigma$ be an orientation preserving diffeomorphism and let $\psi^*g$ be the pullback metric on $\Sigma'$. Then,  
\[ \langle V_{\alpha_1}(x_1)\dots,V_{\alpha_m}(x_m)\rangle_{\Sigma',\psi^*g}  =\langle V_{\alpha_1}(\psi(x_1))\dots,V_{\alpha_m}(\psi(x_m))\rangle_{\Sigma, g}  .\]
\end{enumerate}
\end{proposition}

These results were first established on the sphere \cite{DKRV16}, then extended to all compact surfaces \cite{DRV16_tori,GRV2019}. To fully encode conformal symmetry, one must go beyond Proposition \ref{diffweyl} and use the Segal axioms of CFT in \Cref{sec:segal}. For the moment, we turn to the fundamental example: the three-point function on the sphere.

\section{The DOZZ formula.}\label{sec:dozz}
In the physics literature, the three-point function is the cornerstone of Liouville CFT. Once it is known, all higher correlation functions can in principle be determined by the bootstrap approach. For Liouville theory, the problem reduces   to computing the structure constants of the three-point function on the sphere.

Concretely, we view the Riemann sphere $\mathbb{S}^2$ as the extended complex plane  with the round metric $g_{\mathbb{S}^2}=g(z)|dz|^2$ with $g(z)=4/(1+|z|^2)^2$. Conformal covariance (\Cref{diffweyl}) fixes the form of the three-point function up to a multiplicative constant $C_{\gamma,\mu}(\alpha_1,\alpha_2,\alpha_3)$ called structure constant:
\begin{align}\label{3pointDOZZ}
 \langle  &V_{\alpha_1}(z_1)  V_{\alpha_2}(z_2) V_{\alpha_3}(z_3)  \rangle_{\hat\C,g_{\mathbb{S}^2}}\\
  =&|z_1-z_3|^{2(\Delta_{\alpha_2}-\Delta_{\alpha_1}-\Delta_{\alpha_3})}|z_2-z_3|^{2(\Delta_{\alpha_1}-\Delta_{\alpha_2}-\Delta_{\alpha_3})}|z_1-z_2|^{2(\Delta_{\alpha_3}-\Delta_{\alpha_1}-\Delta_{\alpha_2})}
\Big(\prod_{i=1}^3g(z_i)^{-\Delta_{\alpha_i}}\Big)\nonumber \\
&\times C_0 C_{\gamma,\mu} (\alpha_1,\alpha_2,\alpha_3 ). \nonumber
 \end{align} 
Here    $C_0$ is a universal normalization factor, explicitly
\begin{equation}\label{C_0def}
C_0=\sqrt{\pi} \exp\left(-\tfrac14+2\zeta_R'(-1)-Q^2(1-2\log 2)\right),
\end{equation}
with $\zeta_R$ the Riemann zeta function. 

In a remarkable development, Dorn and Otto   \cite{DornOtto94} and independently Zamolodchikov and Zamolodchikov   \cite{Zamolodchikov96} proposed an explicit formula for the structure constants, now called the DOZZ formula. This proposal was based on consistency arguments in physics and was long regarded as an inspired conjecture. It later received strong support from Teschner \cite{Tesc}, but a full mathematical proof only appeared more than twenty years later in \cite{KRV_DOZZ}.

The formula is expressed in terms of a special holomorphic function introduced by Zamolodchikov. Set $l(z)=\frac{\Gamma (z)}{\Gamma (1-z)}$ with   $\Gamma$   the standard Gamma function.

For $0<\Re(z)<Q$, the function $\Upsilon_{\gamma/2}(z)$ is defined by
\begin{equation}\label{def:upsilon}
\ln \Upsilon_{\frac{\gamma}{2}} (z)  := \int_{0}^\infty  \left ( \Big (\frac{Q}{2}-z \Big )^2  e^{-t}-  \frac{( \sinh( (\frac{Q}{2}-z )\frac{t}{2}  )   )^2}{\sinh (\frac{t \gamma}{4}) \sinh( \frac{t}{\gamma} )}    \right ) \frac{dt}{t}.
\end{equation}
This expression admits analytic continuation to the entire complex plane $\C$, and satisfies the shift relations
\begin{equation}\label{shiftUpsilon}
\begin{aligned}
\Upsilon_{\tfrac{\gamma}{2}} (z+\tfrac{\gamma}{2}) &= l(\tfrac{\gamma}{2}z)(\tfrac{\gamma}{2})^{1-\gamma z}
\Upsilon_{\tfrac{\gamma}{2}} (z) ,\qquad 
\Upsilon_{\tfrac{\gamma}{2}} (z+\tfrac{2}{\gamma})  = l(\tfrac{2}{\gamma}z)
(\tfrac{\gamma}{2})^{\tfrac{4}{\gamma}z-1}
\Upsilon_{\tfrac{\gamma}{2}} (z).
\end{aligned}
\end{equation}
The function $\Upsilon_{\gamma/2}$ has no poles; its zeros are simple (for irrational $\gamma^2$) and form the lattice
$$(-\frac{\gamma}{2} \N-\frac{2}{\gamma} \N) \cup (Q+\frac{\gamma}{2} \N+\frac{2}{\gamma} \N ).$$

With these ingredients, the DOZZ formula for $\alpha_1,\alpha_2,\alpha_3 \in \C$ is
\begin{equation}\label{theDOZZformula}
C_{\gamma,\mu}^{{\rm DOZZ}} (\alpha_1,\alpha_2,\alpha_3 )
= \big(\pi \mu, l(\tfrac{\gamma^2}{4}), (\tfrac{\gamma}{2})^{2 -\gamma^2/2}\big)^{\tfrac{2 Q -\bar{\alpha}}{\gamma}}
\frac{\Upsilon_{\gamma/2}'(0),\Upsilon_{\gamma/2}(\alpha_1),\Upsilon_{\gamma/2}(\alpha_2),\Upsilon_{\gamma/2}(\alpha_3)}
{\Upsilon_{\gamma/2}(\tfrac{\bar{\alpha}}{2}-Q),
\Upsilon_{\gamma/2}(\tfrac{\bar{\alpha}}{2}-\alpha_1),
\Upsilon_{\gamma/2}(\tfrac{\bar{\alpha}}{2}-\alpha_2),
\Upsilon_{\gamma/2}(\tfrac{\bar{\alpha}}{2}-\alpha_3)} ,
\end{equation}
where $\bar{\alpha}=\alpha_1+\alpha_2+\alpha_3$. The formula is meromorphic, with poles precisely at the zeros of the denominator.

The probabilistic construction of Liouville theory eventually confirmed this conjecture:

\begin{theorem}[\cite{KRV_DOZZ}]\label{DOZZth}
Assume the Seiberg bounds hold. Then the structure constant of Liouville theory coincides with the DOZZ formula:
$$C_{\gamma,\mu}(\alpha_1,\alpha_2,\alpha_3 )=C_{\gamma,\mu}^{{\rm DOZZ}} (\alpha_1,\alpha_2,\alpha_3 )$$
Equivalently,
 \begin{equation*}
 \langle  V_{\alpha_1}(0)  V_{\alpha_2}(1) V_{\alpha_3}(\infty)  \rangle_{\mathbb{S}^2,g_{\mathbb{S}^2}} =\frac{C_04^{-\Delta_{\alpha_3}-\Delta_{\alpha_1}}}{2}C_{\gamma,\mu}^{{\rm DOZZ}} (\alpha_1,\alpha_2,\alpha_3 ).
  \end{equation*} 
  \end{theorem}

For further background on both the heuristic derivation and the rigorous proof, we refer the reader to \cite{Vargas}. A key ingredient in the proof is the family of BPZ equations. These arise by inserting a degenerate field, a vertex operator with weight $\alpha \in \{-\tfrac{\gamma}{2}, \tfrac{2}{\gamma}\}$, into a four-point correlation function, which leads to non-trivial shift relations for the three-point function. This strategy, originating in the physics literature \cite{BPZ84},  was initiated in a rigorous probabilistic framework in \cite{KRV19_local}, inspired by  Teschner's argument  \cite{Tesc}. The same method was later used in \cite{Remy20} to establish the celebrated Fyodorov-Bouchaud formula, an important identity in the study of disordered systems \cite{Fyodorov_2008} describing the law of the GMC on the circle.

\section{Segal's axiomatics and amplitudes.}\label{sec:segal}
To fully capture the conformal symmetry of the Liouville model, it is necessary to construct a representation of the symmetry algebra acting on a Hilbert space.  Bridging the gap between the path integral and the
Hilbert space picture has been a central challenge in constructive field theory since the 1970s.  In the 1980s, Segal proposed an axiomatic definition of conformal field theory \cite{Segal87}, designed precisely to encode the symmetry algebra in a geometric way, inspired by the path integral. The main difficulty has long been to exhibit concrete, non-trivial examples satisfying these axioms.  As Segal himself wrote: ``\textit{The manuscript that follows was written fifteen years ago. I simply wanted to justify my definition [...] by checking that all the known examples of conformal field theories did fit the definition. This task held me up}''. Remarkably, the probabilistic formulation of the Liouville model turns out to be particularly well suited to this axiomatic framework.

 \subsection{General philosophy.}
 A field on a manifold $\Sigma$  is a map $\phi:\Sigma\to V$ where $V$ is another manifold, for instance $\phi$ is a section of a fiber bundle over $\Sigma$. In Liouville CFT one takes $V=\R$ and $\Sigma$ a Riemann surface equipped with a metric $g$. The space of fields  $E(\Sigma)$ is typically a function space endowed with a natural topology. For instance, in  Liouville theory,  one may consider the Sobolev space $H^s(\Sigma)$ for some $s<0$.
An action is  a continuous functional 
\[ S_\Sigma: E(\Sigma) \mapsto \C,\]
specifying the local interaction rules of the field. A fundamental feature of such actions is {\it locality}, meaning that $S_\Sigma(\phi)$ can be expressed as an integral
 $S_\Sigma(\phi)=\int_\Sigma s(x,\phi) {\rm v}_g(\dd x)$ 
where $s(x,\phi)$ depends on $\phi$ only through finitely many derivatives at $x$. This immediately implies an additivity property: if $\Sigma=\Sigma_1\cup \Sigma_2$ is decomposed along a codimension-one submanifold $\partial\Sigma_1=\partial\Sigma_2=\Sigma_1\cap\Sigma_2$, then
\begin{align}\label{cutting}
S_\Sigma(\phi)=S_{\Sigma_1}(\phi)+S_{\Sigma_2}(\phi).
\end{align}

In {\it classical field theory}, the main problem is to determine the critical points of $S_\Sigma$, leading to classical variational problems. The canonical example is the Dirichlet energy on a Riemannian surface $(\Sigma,g)$ with boundary, for boundary data $f\in H^{1/2}(\partial\Sigma)$:
\[ S_{\Sigma,f}: E_f:=\{\phi \in H^1(\Sigma)\,|\, \phi|_{\partial \Sigma}=f\} \to \R ,\quad S_{\Sigma,f}(\phi)=\int_{\Sigma}|d\phi|_g^2{\rm dv}_g,\] 
defined on fields $\phi\in H^1(\Sigma)$ with $\phi|_{\partial\Sigma}=f$. The unique minimizer is the harmonic extension $\phi_0$, solving $\Delta_g\phi_0=0$ with boundary condition $f$. 

In   {\it Euclidean Quantum  Field Theory}, one replaces minimization by integration and formally considers the measure
\begin{equation} \label{PIsegal}
e^{-S_\Sigma(\phi)}D\phi 
\end{equation}
over the space of fields $E(\Sigma)$  
where $D\phi$ is the heuristic ``Lebesgue measure'' on $E(\Sigma)$, i.e. some measure invariant under the symmetries of the action functional.  Making sense of \eqref{PIsegal} rigorously is subtle and often impossible, but the expression is an invaluable guide. In particular, combining \eqref{cutting} with the idea that $D\phi$ should factorize accordingly  leads to a natural gluing property for path integrals. Concretely, for an observable $F(\phi)=F_1(\phi_{|\Sigma_1})F_2(\phi_{|\Sigma_2})$ adapted to a decomposition $\Sigma=\Sigma_1\cup\Sigma_2$, one is led to the relation
 $$\int_{E(\Sigma)} F(\phi)e^{-S_\Sigma(\phi)}D\phi = \int_{E(\Sigma_1\cap\Sigma_2)} \mc{A}_{\Sigma_1,F_1}(\varphi)\mc{A}_{\Sigma_2,F_2}(\varphi)D\varphi$$
where  the conditional path integral 
\begin{equation}\label{formamp}
 \mc{A}_{\Sigma_i,F_i}(\varphi):= 
 \int_{E(\Sigma_i); \phi|_{\partial\Sigma_i}=\varphi} F_i(\phi)e^{-S_{\Sigma_i}(\phi)}D\phi
 \end{equation}
is the {\it amplitude} of the manifold  $\Sigma_i$ with boundary  $\partial\Sigma_i$, and boundary field $\varphi$. From a probabilistic perspective, this factorization is nothing but a Markov property: conditioning the field $\phi$ on its boundary values along $\Sigma_1\cap\Sigma_2$ makes the two sides independent. This interpretation is particularly natural for the Gaussian Free Field, which indeed satisfies such a Markov property (see e.g. \cite{Sheffield07,Berestycki17}).  
 
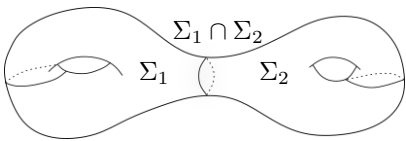
\begin{figure}
\begin{center}
   \begin{tikzpicture}
    \node[inner sep=0pt] (pant) at (0,0)
{\includegraphics[scale=0.4]{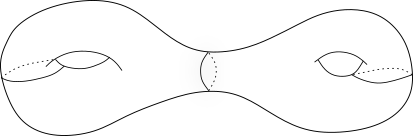}};
    \draw (-0.55,0.55) node[right,black]{$\Sigma_1\cap\Sigma_2$} ;
     \draw (-1,0) node[right,black]{$\Sigma_1$} ;
       \draw (0.6,0) node[right,black]{$\Sigma_2$} ;
\end{tikzpicture}
\caption{A surface cut along a circle $\Sigma_1\cap\Sigma_2$}
\label{picsegal}
\end{center}
\end{figure}

 
Graeme Segal's  axiomatization of  CFT  captures locality and factorization via amplitudes assigned to Riemann surfaces with boundary. These axioms imply that a CFT is determined by its values on   fundamental geometric building blocks: disks with 2 marked points, annuli with 1 marked point, and pairs of pants. Broader categorical frameworks have since emerged; for instance, Atiyah's axioms for topological QFTs \cite{Atiyah}, Functorial Quantum  Field Theories \cite{Lurie}, and the homotopical formulation via Costello-Gwilliam's factorization algebras \cite{CostelloGwilliam}.
 
 \subsection{Segal's formalism in Liouville theory} 
 
 In Segal's axiomatic formulation of conformal field theory, the basic data consist of a Hilbert space $\mc{H}$ together with a collection of amplitudes. To each Riemann surface $\Sigma$ with analytically parametrized boundary components and equipped with a metric $g$, one assigns a linear operator. The surface may also carry marked points ${\bf x}=(x_1,\dots,x_m)$ with weights $\boldsymbol{\alpha}=(\alpha_1,\dots,\alpha_m)$ attached. Each boundary component is tagged as either {\it In} or {\it Out}, depending on whether the  orientation induced by the parametrization disagrees or agrees with that of $\Sigma$. A copy of $\mc{H}$ is attached to each boundary component, so that the amplitude $ \mc{A}_{\Sigma,g,{\bf x},\bs{\alpha}}$
 is realized as a linear operator $\mc{H}^{\otimes {\rm In}} \to \mc{H}^{\otimes {\rm Out}}$, where In-Out stand here for the number of In-Out boundary components. The total number of boundary components is $b:={\rm In}+{\rm Out}$. In addition, amplitudes are required to be Hilbert-Schmidt.  The Hilbert space is typically a functional space of maps defined on the circle: it encodes the restriction of the field to a given boundary component.  
 
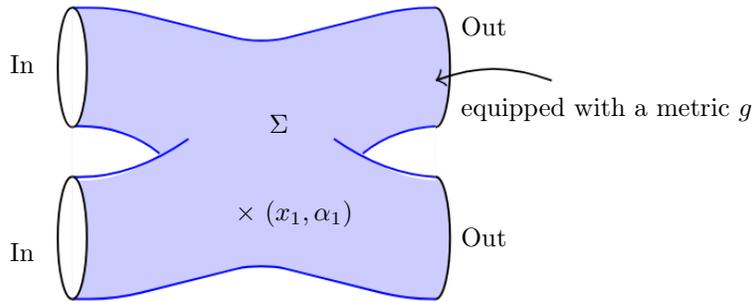
\begin{figure}[h]
\begin{center}
   \begin{tikzpicture}[]
   \node (pant) at (0,0) {\includegraphics[width=0.4\columnwidth]{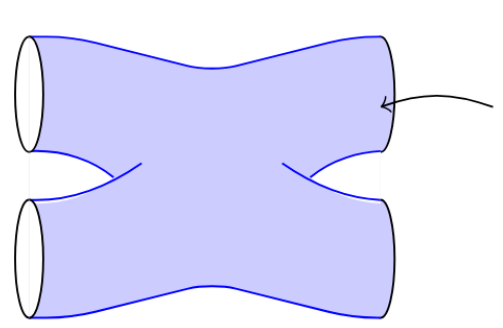}};
    \draw (-0.55,0.2) node[right,black]{$\Sigma$} ;
     \draw (-4,1) node[right,black]{${\rm In}$} ;
      \draw (-4,-1.5) node[right,black]{${\rm In}$} ;
       \draw (2,1.5) node[right,black]{${\rm Out}$} ;
        \draw (2,-1.3) node[right,black]{${\rm Out}$} ;
         \draw (-1,-1) node[right,black]{$\times$  $(x_1,\alpha_1)$} ;
         \draw (2,0.4) node[right,black]{equipped with a metric $g$} ;
\end{tikzpicture}
\caption{Surface with In/Out boundary components and 1 marked point $x_1$ with weight $\alpha_1$}
\label{picampl}
\end{center}
\end{figure}

Formally, $\mc{A}_{\Sigma,g,{\bf x},\bs{\alpha}}$ should correspond to the path integral
 \begin{equation}\label{formampL}
 \mc{A}_{\Sigma}(\bs{\varphi}):= 
 \int_{E(\Sigma); \phi|_{\partial_i\Sigma}=\varphi_i} \big(\prod_{j=1}^mV_{\alpha_j}(x_j)\big)e^{-S_{\Sigma}(\phi)}D\phi
 \end{equation}
where $\bs{\varphi}=(\varphi_1,\dots,\varphi_b)$ denotes the boundary fields along the $b$ boundary components $\partial \Sigma_i$, and $V_{\alpha_j}$ are vertex operators. The Segal axioms postulate, in particular, natural transformation rules (cf. \Cref{diffweyl}) and a gluing property: amplitudes behave functorially under sewing of surfaces. More precisely, if $(\Sigma^1,g^1,{\bf x}^1,\bs{\alpha}^1)$ and $(\Sigma^2,g^2,{\bf x}^2,\bs{\alpha}^2)$ are two such surfaces, if  $(\Sigma,g,{\bf x},\bs{\alpha})$ is obtained by gluing an {\rm Out}-boundary component of $\Sigma^1$ with an {\rm In}-boundary component of $\Sigma^2$  according to their boundary parametrizations (with  ${\bf x}={\bf x}^1\cup {\bf x}^2$ and $\bs{\alpha}=(\bs{\alpha}^1,\bs{\alpha}^2)$), and   the two metrics $g_1$ and $g_2$ glue nicely\footnote{\cite{GKRV21_Segal} requires the metrics to be admissible near the boundary components but we will not elaborate more on this.} so as to form the metric $g$ on $\Sigma$, then one has
\begin{equation}\label{partial_trace}
\mc{A}_{\Sigma,g,{\bf x},\bs{\alpha}}
= 
\underbrace{\mc{A}_{\Sigma^2,g^2,{\bf x}^2,\bs{\alpha}^2}}_{ \mc{H}^{\otimes {\rm In}_2}\to \mc{H}^{\otimes{\rm Out}_2}}   
\circ_{ij} 
\underbrace{\mc{A}_{\Sigma^1,g^1,{\bf x}^1,\bs{\alpha}^1}}_{ \mc{H}^{\otimes {\rm In}_1}\to \mc{H}^{\otimes{\rm Out}_1}} 
\;\;\in \mc{H}^{\otimes ({\rm In}_1+{\rm In}_2-1)} \to \mc{H}^{\otimes ({\rm Out}_1+{\rm Out}_2-1)}.
\end{equation}
Here $\circ_{ij}$ denotes composition of the $i$-th Out-component of $\mc{A}_{\Sigma^1}$ with the $j$-th In-component of $\mc{A}_{\Sigma^2}$.  

Because the Liouville action \eqref{intro:action} is local, one expects Segal's axioms to hold. This was indeed proved in \cite{GKRV21_Segal}. Of course, the construction involves delicate choices: the Hilbert space $\mc{H}$ is not unique, and the formal ``Lebesgue measure on maps'' is ill-defined. A natural choice arises from the Gaussian free field. Any real-valued function $\varphi$ on the unit circle can be expanded as the trigonometric series
\begin{equation}\label{circleGFF}
\varphi(\theta)=c+\sum_{n\neq 0}\varphi_n e^{in\theta},\qquad \theta\in \R,
\end{equation}
with Fourier coefficients obeying $\varphi_{-n}=\overline{\varphi_n}$. Reparametrizing $\varphi_n$ for $n>0$ as $\varphi_n=\tfrac{x_n+iy_n}{2\sqrt{n}}$, one obtains the Gaussian measure
\begin{equation}\label{mu0}
\mu_0=\dd c \bigotimes_{n=1}^\infty \frac{e^{-\tfrac{x_n^2}{2}-\tfrac{y_n^2}{2}}}{2\pi}\,\dd x_n\dd y_n,
\end{equation}
which has infinite total mass due to the zero mode. In \cite{GKRV21_Segal}, the Hilbert space is chosen to be $L^2(\mu_0)$. Under $\mu_0$, the random series \eqref{circleGFF} coincides with the restriction of the whole-plane GFF to the unit circle.  

From this perspective, the probabilistic construction of Liouville amplitudes \eqref{formampL} mirrors the path integral approach in \Cref{sec:proba}, with one modification: the GFF on $\Sigma$ is now conditioned to match the prescribed boundary values $\bs{\varphi}$ on the boundary components. This variant can be described  as a Dirichlet GFF plus the harmonic extension of the boundary fields. 

\begin{theorem}[Informal]\label{th:segal}
If $\alpha_j<Q$ for $j=1,\dots,m$ and $\sum_j\alpha_j>\chi(\Sigma) Q$ then the Liouville amplitudes obey the Segal axioms.
\end{theorem}

Segal's axioms provide a powerful geometric and functorial framework: amplitudes are defined directly from  the Riemannian structure of Riemann surfaces with boundary. It will be convenient to reformulate these rules in the {\it operator formalism}, where the emphasis shifts from surfaces to the Hilbert space itself. In this picture, boundary circles are interpreted as state spaces, amplitudes become operators acting on $\mc{H}$, and the dynamics of the theory can be encoded in spectral data such as the Hamiltonian and the symmetry algebra. This Hilbert space viewpoint is closer to the original language of physics and will serve as a bridge between the probabilistic path integral picture and the algebraic structure underlying conformal field theory.

\section{Hilbert space picture and operator formalism.}\label{sec:annulus}
In the operator formalism, the central object is the Hilbert space $\mc{H}$ itself, whose vectors represent states of the theory and whose operators implement the action of observables and symmetries. Correlation functions are then reconstructed from matrix elements of such operators, while the conformal symmetry manifests through the representation of the Virasoro algebra acting on $\mc{H}$. 

Segal's amplitudes provide a way to implement these ideas concretely. Let $\mc{S}$ denote the class of Riemann surfaces with two parametrized boundary components, one incoming and one outgoing, with the topology of an annulus. They form a complex Lie semigroup under gluing of their boundary, and the amplitudes associated to $\mathcal{S}$ form a projective representation of $\mathcal{S}$ by operators   on $\mc{H}$. Because of the metric dependence (Weyl anomaly), this representation is not genuine but  projective, with the anomaly measured by the central charge. Differentiating this projective representation produces a representation of the central extension of the Lie algebra of $\mathcal{S}$, namely the Virasoro algebra.

To motivate this construction, Segal  \cite{Segal87}, and independently Neretin  \cite{Ner87,Ner1990}, observed that the group ${\rm Diff}^+(\mathbb{S}^1)$ of orientation-preserving   diffeomorphisms of the circle is a real infinite-dimensional Lie group without a genuine complexification. Nevertheless, the semigroup $\mc{S}$ can reasonably  be viewed as a sub-semigroup of the (non-existent) complexification of ${\rm Diff}^+(\mathbb{S}^1)$. Since the Lie algebra of ${\rm Diff}^+(\mathbb{S}^1)$ is the Witt algebra of vector fields on the circle, its central extension is the Virasoro algebra. This perspective was carried out in the Liouville setting in \cite{BGKRV,BGKR1}, yielding a representation of the Virasoro algebra on $\mc{H}$. This representation encodes the fundamental structural features of the theory, as we will later explain.

%
%

\begin{figure}[h] 
 \begin{center}
\begin{tikzpicture}[scale=0.8]
\draw[fill=lightgray,draw=blue,very thick] (0,0) circle (2) ;
\coordinate (Z1) at (1.5,0) ;
\coordinate (Z2) at (0,0) ;
\coordinate (Z3) at (-0.8,1.3) ;
\coordinate (Z4) at (-0.7,-0.9) ;
\draw[>=latex,draw=red,very thick,fill=white] (Z1) to[out=120,in=40] (Z2) to [out=220,in=40](Z3) to [out=220,in=70](Z4) to [out=250,in=300](Z1);
  \draw (0,-2) node[above,red]{$ f(\T)$} ;
\draw (2,0) node[right,blue]{$ \T$} ;
 \draw (0.5,0.5) node[above]{$\mathbb{A}_f$} ;
 \draw (0.3,-0.9) node[above]{$f(\D)$} ;
\end{tikzpicture}
\end{center}
\caption{The annulus $\mathbb{A}_f$, for $f\in \mc{S}_+$.}\label{pic:af}
\end{figure}
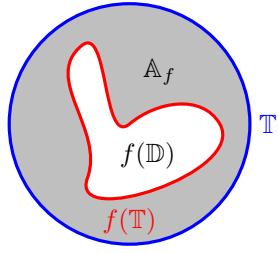

 More precisely, let $\mc{S}_+$ be the semigroup of holomorphic annuli. It consists of biholomorphic maps $f$  on the unit disk $\D$ that are smooth up to the boundary $ \partial\D$, with $f(0)=0$ and $f(\D)\subset \D$. The semigroup is an infinite-dimensional complex manifold, and  its tangent space is  the space ${\rm Hol}^{\bullet}(\D)$ of holomorphic vector fields $v(z)\partial_z$ on $\D$, with $v(0)=0$. For such an $f\in \mathcal{S}_+$, one can consider the annulus $\mathbb{A}_f$ lying between the unit circle $ \partial\D$ and $f( \partial\D)$, see \Cref{pic:af}, with the boundary parametrizations 
 $$\zeta_1: \partial\D\to \partial\D, \quad \zeta(e^{i\theta}):=e^{i\theta} ,\quad  \zeta_2: \partial\D\to f(\partial\D), 
\quad \zeta_2(e^{i\theta}):=f(e^{i\theta}). $$

The amplitudes of such annuli, equipped with a metric $g$, define a projective semigroup of operators $\mc{A}_{\mathbb{A}_f,g}:\mc{H}\to\mc{H}$. In fact, up to removing an explicit multiplicative factor depending on $f$ and $g$, \cite{BGKR1} constructs a semigroup $\mathbf{T}_f:=e^{-W(f,g)}\mc{A}_{\mathbb{A}_f,g}$ for some $W(f,g)\in\R$, which admits a probabilist Feynman-Kac type representation. Furthermore the map $f\in \mc{S}_+\mapsto \mathbf{T}_f$ is differentiable, with 
\begin{equation}\label{varyann}
\forall v\in {\rm Hol}^{\bullet}(\D),\quad D_v\mathbf{T}_f=-\mathbf{T}_f\mathbf{H}_w, \qquad \text{with}\quad  \mathbf{H}_w=\sum_{n\geq 0}w_n\mathbf{L}_n+\bar{w}_n\widetilde{\mathbf{L}}_n 
\end{equation}
where the coefficients $(w_n)$ arise from the Taylor expansion at $0$ of $w(z) = v(z)/f'(z)$: $w(z) = -\sum_{n\ge 0} w_n z^{n+1}$. 
Here $(\mathbf{L}_n)_{n\ge 0}$ and $(\widetilde{\mathbf{L}}_n)_{n\ge 0}$ are unbounded operators on $\mc{H}$, which can be extended to $n\in \mathbb{Z}$ by $\mathbf{L}_{-n}=\mathbf{L}_n^*$ and  $\widetilde{\mathbf{L}}_{-n}=\widetilde{\mathbf{L}}_n^*$ for $n\geq 0$. These families form two commuting representations of the Virasoro algebra, with central charge $c_{\rm L}$:
\begin{align}
  [{\bf L}_n,{\bf L}_{m}]= & (n-m){\bf L}_{n+m} +\frac{c_{\rm L}}{12}(n^3-n)\delta_{n,-m}, \qquad \text{and similarly for } (\widetilde{\mathbf{L}}_n)_{n\geq 0}\\
    [{\bf L}_n,\widetilde{\mathbf{L}}_{m}]=&0.
\end{align}
A more direct construction of $\mathbf{L}_n$ and $\widetilde{\mathbf{L}}_n$ for $n<0$ uses annuli with  general outer boundaries, but the analysis is entirely analogous.

The Hamiltonian of the Liouville model is defined as  
$$
 \mathbf{H}:=\mathbf{L}_0+\widetilde{\mathbf{L}}_0.
 $$
It plays a role reminiscent of the quadratic Casimir operator for representations of semisimple Lie algebras, in the sense that it governs the classification of irreducible Virasoro representations appearing in the theory. Unlike the Casimir, however, $\mathbf{H}$ does not commute with the full Virasoro algebra. Geometrically, $\mathbf{H}$ corresponds to the generator of dilations. Indeed, consider the holomorphic contraction $f(z) = r z$ with $r \in (0,1)$, and differentiate in the direction of the vector field $v(z)\partial_z = -z\partial_z$. One finds
$$
D_v\mathbf{T}_f=-\frac{1}{r}\mathbf{T}_f\mathbf{H}.
$$
Thus the flow generated by $v(z)\partial_z$ acts on $\mc{S}_+$ precisely by radial dilations, and the Hamiltonian implements this symmetry at the level of the Hilbert space.

\section{Spectrum of the Hamiltonian}

We now turn to the spectral analysis of the Hamiltonian. In the coordinate system of \cref{circleGFF}, it admits the explicit expression
\begin{equation}
\mathbf{H}=\mathbf{H}_0+\mu e^{\gamma c} V(\tilde{\varphi}), \qquad  \mathbf{H}_0:=-\tfrac{1}{2}\partial^2_c+\tfrac{Q^2}{2}+\sum_{n\geq 1}n(\partial_{x_n}^*\partial_{x_n}+\partial_{y_n}^*\partial_{y_n}) 
\end{equation}
where the adjoints are taken with respect to the measure $\mu_0$ in \cref{mu0}.
We have decomposed the field $\varphi$ as $\varphi=c+\tilde{\varphi}$ to single out the zero mode, and the potential is expressed in terms of a Gaussian multiplicative chaos on the circle:
\begin{equation}
V(\tilde{\varphi}):=\int_0^{2\pi}e^{\gamma \tilde{\varphi}(\theta)-\frac{\gamma^2}{2}\E[\tilde{\varphi}^2(\theta)]}\,\dd \theta.
\end{equation}
Thus, $\mathbf{H}_0$ describes the free theory and has the structure of an infinite-dimensional Laplacian, while the interaction $\mu e^{\gamma c}V(\tilde{\varphi})$ acts by multiplication. In this way, $\mathbf{H}$ naturally appears as a Schr\"odinger-type operator.

 In physics, the spectral picture was first discussed in physics by Curtright and Thorn \cite{PhysRevLett.48.1309}, and later revisited by Teschner \cite{Teschner_revisited}, who emphasized its scattering interpretation.
 
 From the mathematical side, the operator without the $c$-variable was studied in \cite{HK}, where it was shown to be essentially self-adjoint for $\gamma\in(0,1)$, the regime in which the potential is well-defined as an $L^2$ random variable. For $\gamma\in [1,\sqrt{2})$, the potential only belongs to $L^{1+\epsilon}$, and when $\gamma\geq \sqrt{2}$ it no longer exists as a random variable, but only as a random measure. The complete spectral analysis of the full operator was   carried out in \cite{GKRV20_bootstrap}, relying on scattering theory. The key observation is that the potential $e^{\gamma c}V$ acts as a barrier as $c\to+\infty$, while it vanishes as $c\to-\infty$. Hence, eigenstates of $\mathbf{H}$ can be reconstructed from their asymptotics at $c\to-\infty$, where they approximate the eigenstates of $\mathbf{H}_0$, which are explicitly computable.

 \begin{figure}[h]
 \begin{center}
\begin{tikzpicture}
  \draw[->] (-3, 0) -- (2, 0) node[right] {$c$};
  \draw[->] (0,0) -- (0, 2.5) node[above] {$$};
  \draw[scale=0.8, domain=-3.5:1.3, smooth, variable=\x, black] plot ({\x}, {exp(\x)});
  \draw (0.8,2) node[right,black]{$e^{\gamma c}$} ;
   \draw[->,draw=red,very thick] (-3,2.5) --node[midway,below]{$\textrm{incoming wave}$} (-1,2.5) ;
    \draw[->,draw=blue,very thick] (-1,1.5) --node[midway,below]{$\textrm{reflected wave}$} (-3,1.5) ;
\end{tikzpicture}
\caption{The toy model potential $V=e^{\gamma c}$}
 \end{center}
\end{figure}
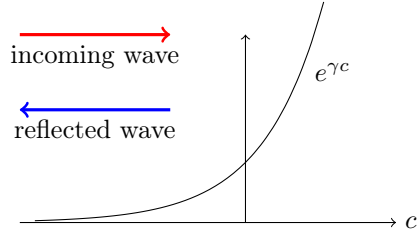
A useful toy model is given by the operator
$${\bf H}_{\rm mod}:=-\frac{1}{2}\partial^2_c+\tfrac{Q^2}{2}+e^{\gamma c}\qquad \text{on} \quad L^2(\R).$$
Its spectrum is purely continuous, equal to $[\tfrac{Q^2}{2},\infty)$, and for each $p\in\R$ there exists a unique solution $\psi_p$ of
\begin{equation}
({\bf H}_{\rm mod}-2\Delta_{Q+ip})\psi_{p}(c)=0, \qquad \Delta_{Q+ip}=\frac{Q^2+p^2}{4}
\end{equation}
 which admits the scattering form   
\[ \psi_p|_{\R^+}\in L^2(\R^+), \quad  \psi_p|_{\R^-}=e^{ipc}+R(p)e^{-ipc}+\tilde{\psi}_p, \,\textrm{with } \tilde{\psi}_p\in L^2(\R^-).\]
Here $e^{ipc}$ represents an incoming plane wave, while $R(p)e^{-ipc}$ is the reflected outgoing wave as $c\to-\infty$. The functions $\psi_p$ can be expressed in terms of modified Bessel functions and thus extend meromorphically in $p\in\C$ as smooth functions of $c$.

This scattering picture extends in large part to the full Liouville Hamiltonian.  One finds that $\mathbf{H}$ is a positive self-adjoint operator with continuous spectrum
 \begin{equation}
{\rm Spec}(\mathbf{H})=[\tfrac{Q^2}{2},+\infty[.
\end{equation}

To describe the eigenstates of ${\bf H}$, we need to introduce   $\mc{T}$   the set of Young diagrams, that is, finite sequences of decreasing nonnegative integers.
For $\nu\in\mc{T}$ we write $|\nu|$ for its length, i.e. the sum of the integers in the sequence.
The Schapovalov form (also called the Virasoro Gram matrix) is a matrix $(F_\alpha(\nu,\nu'))_{|\nu|=|\nu'|=N}$ indexed by Young diagrams of length $N$, with parameter $\alpha\in\C$.
It naturally encodes the structure of the Virasoro Verma module, see  \cite[Section 4.4]{GKRV20_bootstrap} for the precise definition and main properties.
This matrix is invertible outside the Kac table, i.e.
 $$
 \alpha\not\in \{{\alpha_{r,s}} \mid \,\,r,s\in \N^\ast ,rs \leq \max(|\nu|, |\tilde \nu |)\}\quad \quad\text{with }\quad {\alpha_{r,s}}=Q-r\frac{\gamma}{2}-s\frac{2}{\gamma}.
 $$

The main spectral theorem of \cite{GKRV20_bootstrap} can be summarized as follows.
 \begin{theorem} \label{th:desc:liouv}
For each $\nu,\tilde{\nu}\in\mc{T}$ with $|\nu|+|\tilde\nu|=N$, there is $C_N>0$ and a connected open set $\Omega_{N}\subset \C$ containing
$(Q+i\R)\cup (-\infty,Q-C_N)$, and an 
analytic family 
\begin{equation}
\alpha \in \Omega_{\nu,\tilde{\nu}} \mapsto \Psi_{\alpha,\nu,\tilde{\nu}}\in e^{(-|Q-{\rm Re}(\alpha)|+\epsilon)c\mathbf{1}_{c<0}}\mc{H}
\end{equation}
 such that
\begin{equation}
({\bf H}-2\Delta_{\alpha}-|\nu|-|\tilde\nu|)\Psi_{\alpha,\nu,\tilde{\nu}}=0.
\end{equation} 
Specializing to  the spectrum line $\alpha\in Q+i\R$ the following spectral decomposition holds: for $u,v\in\mc{H}$  
\begin{equation}\label{spectral}
\langle u,v\rangle_{\mc{H}}=\frac{1}{2 \pi}\sum_{\nu,\nu',\tilde{\nu},\tilde{\nu}'} \int_{\R_+} \langle u,
\Psi_{Q+ip,\nu',\tilde{\nu}'} \rangle_{\mc{H}} \langle \Psi_{Q+ip,\nu,\tilde{\nu}}, v\rangle _{\mc{H}}F^{-1}_{Q+ip}(\nu,\nu')F^{-1}_{Q+ip}(\tilde\nu,\tilde\nu')\, \dd p 
\end{equation}
where the sum runs over Young diagrams $\nu,\tilde\nu,\nu',\tilde\nu'\in \mc{T}$ such that $|\nu|=|\nu'|$ and $|\tilde\nu|=|\tilde\nu'|$. 
\end{theorem}

In this framework, the notion of eigenstate must be understood in a generalized sense: these functions do not belong to the Hilbert space itself, but rather to suitable weighted extensions of it. This is completely analogous to plane waves $c\mapsto e^{\alpha c}$, which serve as generalized eigenstates of the Laplacian on $\R$. The family $(\Psi_{Q+ip,\nu,\tilde{\nu}})_{p\in\R,\nu,\tilde{\nu}\in\mc{T}}$ forms a complete set of generalized eigenstates, but not an orthogonal one. The Schapovalov form thus plays the role of Gram-Schmidt coefficients for these states.

The theorem above is essentially a consequence of general scattering theory and, by itself, does not reveal much about the model-dependent structure of the eigenstates. For this purpose, the probabilistic construction becomes crucial, and Segal's functor provides an efficient way to guess the form of eigenstates. Indeed, the propagator $e^{-t\mathbf{H}}$ corresponds to the amplitude of an annulus. Gluing a straight annulus to a disk with a marked point at the center leaves the geometry unchanged (a disk with one marked point at the center). Translated via Segal's functor, this shows that disk amplitudes must be eigenstates of the Hamiltonian.  Concretely, the disk amplitude on the unit disk with a marked center of weight $\alpha<Q$ is given by
\begin{equation}\label{probaES}
\Psi_{\alpha}(\varphi)=e^{(\alpha-Q)c}\E [ e^{-\mu \int_{\D}|x|^{-\gamma \alpha}e^{\gamma P\varphi(x) }M^\D_\gamma(dx)}] 
\end{equation}
where $P\varphi$ denotes the harmonic extension of the boundary field $\varphi$, and $M^\D_\gamma$ is the GMC measure on $\D$ built from the Dirichlet GFF $X_{\D}$, , with expectation taken over this field.  This expression is analytic in $\alpha$ in a neighborhood of the half-line $\alpha<Q$, and must therefore coincide with the analytic continuation of the generalized eigenstate $ \Psi_{\alpha,\emptyset,\emptyset}$. This analytic bridge between the probabilistic region ($\alpha<Q$) and the spectral line ($\alpha\in Q+i\R$) allows one to transfer probabilistic computations to the spectral side. The same idea extends to descendant states $\Psi_{\alpha,\nu,\tilde{\nu}}$ for nontrivial Young diagrams. Using this approach, it was shown in \cite{BGKRV} that
\begin{proposition}\label{prop:mainvir1}  
 For $p\in\R_+$, $\nu,\tilde{\nu}\in\mc{T}$ one has 
\begin{equation}\label{viraction}
\Psi_{Q+ip,\nu,\tilde{\nu}} =\mathbf{L}_{-\nu}\tilde{\mathbf{L}}_{-\tilde\nu}  \: \Psi_{Q+ip}
\end{equation}
 where, for $\nu=(m_1,\dots,m_k)\in\mathcal{T}$, $\mathbf{L}_{-\nu}$ stands for the composition of operators $\mathbf{L}_{-\nu}:=\mathbf{L}_{-m_1}\dots \mathbf{L}_{-m_k}$, and similarly for $\tilde{\mathbf{L}}_{-\nu}  $. Moreover, $\Psi_{\alpha,\nu,\tilde{\nu}}$ admit an analytic extension to $\alpha \in \C$ in suitable weighted  spaces and the relation above extends as well.
 \end{proposition}

A further striking feature is the duality relation \cite{BGKRV}, for all $\alpha\in \C \setminus (Q\pm (\frac{2}{\gamma}\N_0+\frac{\gamma}{2}\N_0))$  
\[ \Psi_{2Q-\alpha,\nu,\tilde{\nu}}=R(2Q-\alpha)\Psi_{\alpha,\nu,\tilde{\nu}}\]
with $R(\alpha)$ the scattering (or reflection) coefficient  
\begin{equation}\label{reflection}
R(\alpha):=-\Big(\pi \mu \frac{\Gamma(\frac{\gamma^2}{4})}{\Gamma(1-\frac{\gamma^2}{4})}\Big)^{2\frac{(Q-\alpha)}{\gamma}}\frac{\Gamma(-\frac{\gamma(Q-\alpha)}{2})\Gamma(-\frac{2(Q-\alpha)}{\gamma})}{\Gamma(\frac{\gamma(Q-\alpha)}{2})\Gamma(\frac{2(Q-\alpha)}{\gamma})}.
\end{equation}
The existence of such a coefficient is expected from scattering theory, but its explicit identification is a genuine achievement of the probabilistic approach. Indeed,  scattering theory yields the following expansion for the eigenstates at $c\to-\infty$
\begin{equation}\label{scatt_asymp}
 \Psi_{Q+iP,\nu,\tilde{\nu}}= \Psi^0_{Q+iP,\nu,\tilde{\nu}}+R(Q+iP) \Psi^0_{Q-iP,\nu,\tilde{\nu}}+G_{Q+iP}
\end{equation}
where $G_{Q+iP}\in \mc{H}$ is a correction term, and  $(\Psi^0_{Q+iP,\nu,\tilde{\nu}})_{p\in\R,\nu,\tilde{\nu}\in\mc{T}}$ stand for the eigenstates of the free Hamiltonian $\mathbf{H}_0$, obeying the same relation as \cref{viraction} for the Virasoro generator of the free theory. Analyzing the expansion the probabilistic expression \cref{probaES} of $\Psi_\alpha$ as $c\to-\infty$ and comparing with \cref{scatt_asymp} yields a probabilistic formula for $R(\alpha)$, developed  in \cite{RhodesVargas2019}, and further expanded in \cite{Wong2020}.  The same probabilistic expression has been found in \cite{KRV_DOZZ} as the limit of the structure constants
\begin{equation}
4R(\alpha)=\lim_{\epsilon\to 0}\epsilon C_{\gamma,\mu}(\alpha,\epsilon,\alpha),
\end{equation}
which shows that the scattering coefficient coincides with the two-point function of Liouville CFT. Interestingly,  the probabilistic representation reveals another striking feature: $R(\alpha)$ naturally appears as the partition function of the quantum sphere introduced in \cite{MatingOfTrees}. Through the DOZZ formula, this gives the explicit expression \eqref{reflection}.

\section{Conformal bootstrap.} \label{sec:bootstrap}
From Segal's axiomatic perspective, a correlation function on a Riemann surface is constructed by assigning amplitudes to pairs of pants and then gluing them along their parametrized boundaries. In this setting, a ``pair of pants'' is understood in the generalized sense that marked points count as ends: thus one may have a disk with two marked points, an annulus with one marked point, or the usual pair of pants with three boundary components. The resulting amplitude can be analyzed by inserting the spectral decomposition of the Hilbert space at each gluing step. This reduces the computation of the correlation function to determining the action of pair-of-pants amplitudes on the eigenstates of the Hamiltonian (i.e. Virasoro modules). In practice, one finds that primaries contribute through the DOZZ structure constants, which determine the three-point functions, while descendants contribute universally through conformal blocks, namely special functions  determined by the representation theory of the Virasoro algebra. Thus, Segal's categorical formalism naturally leads to the bootstrap decomposition of correlation functions into DOZZ constants and conformal blocks.
 
This strategy was carried out in \cite{GKRV21_Segal} for arbitrary correlation functions. To make the construction more concrete, we will focus here on two illustrative cases: the four-point function on the Riemann sphere and the partition function on a genus-two surface.

\bigskip 
\paragraph*{Four point correlation on the Riemann sphere.}

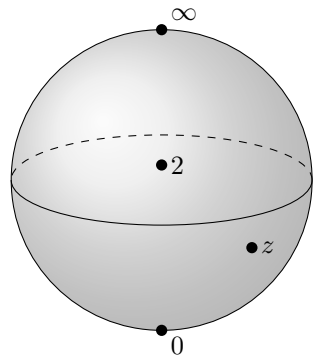
\begin{figure}[h] 
\begin{center}
 \begin{tikzpicture}
  \shade[ball color = gray!40, opacity = 0.4] (0,0) circle (2cm);
  \draw (0,0) circle (2cm);
  \draw (-2,0) arc (180:360:2 and 0.6);
  \draw[dashed] (2,0) arc (0:180:2 and 0.6);
   \node at (1.2,-0.9) [circle,fill,inner sep=1.5pt]{};
    \node at (0,-2) [circle,fill,inner sep=1.5pt]{};
     \node at (0,2) [circle,fill,inner sep=1.5pt]{};
       \node at (0,0.2) [circle,fill,inner sep=1.5pt]{};
     \draw (1.2,-0.9) node[right,black]{$z$} ;
      \draw (0,0.2) node[right,black]{$2$} ;
        \draw (0,-2.2) node[right,black]{$0$} ;
           \draw (0,2.2) node[right,black]{$\infty$} ;
\end{tikzpicture} \caption{The sphere with $4$ points, cut into $2$ disks along the equator.}\label{picsphere2} 
\end{center}
\end{figure}

By M\"obius invariance, the four marked points can be fixed at  
$$ {\bf x}=(x_1,x_2,x_3,x_4)=(0,z,2,\infty),\qquad |z|<1,
$$
with corresponding weights $\bs{\alpha}=(\alpha_1,\alpha_2,\alpha_3,\alpha_4)$ satisfying $\alpha_i<Q$,  
$\alpha_1+\alpha_2>Q$ and $\alpha_3+\alpha_4>Q$.  We view the Riemann sphere as obtained by gluing the unit disk $\D$ and its exterior $\D^*$ along the equator. The metric $g=g(z)|dz|^2$ is chosen so as to be well adapted to this gluing: near the equator it takes the form $|z|^{-2}|dz|^2$, and for convenience we normalize it so that $g(0)=1$ and it is invariant by $z\mapsto 1/z$.  Both disks admit the parametrization $\zeta(e^{i\theta})=e^{i\theta}$  of their boundary (incoming for $\D^*$, outgoing for $\D$). See   \Cref{picsphere2}.  
Finally,   we choose the metric $g$ so as to make it flat near the insertions. 
Segal's gluing property then yields, with ${\bf x}_1=(0,z)$ and ${\bf x}_2=(2,\infty)$, and corresponding weights
  $\bs{\alpha}_1=(\alpha_1,\alpha_2)$ and $\bs{\alpha}_2=(\alpha_3,\alpha_4)$
\begin{equation}\label{split_Sphere}
\langle \prod_{i=1}^4 V_{\alpha_i}(x_i)\rangle_{\mathbb{S}^2,g}=\langle \mc{A}_{\D,g,{\bf x}_1,\bs{\alpha}_1},\mc{A}_{\D^*,g,{\bf x}_2,\bs{\alpha}_2 }\rangle_{\mc{H}}
\end{equation}
where both amplitudes are elements of $\mc{H}$, as ensured by the conditions on the weights (see \Cref{th:segal}).
Inserting the spectral decomposition \cref{spectral} gives
\begin{align*}
&\langle \prod_{i=1}^4 V_{\alpha_i}(x_i)\rangle_{\mathbb{S}^2,g} =\frac{1}{2\pi}\int_0^\infty\sum_{\nu,\nu',\tilde{\nu},\tilde{\nu}'}
\langle \mc{A}_{\D,g,{\bf x}_1,\bs{\alpha}_1 },\Psi_{Q+ip,\nu,\tilde{\nu}}\rangle_{\mc{H}}\langle \Psi_{Q+ip,\nu',\tilde{\nu}'}, \mc{A}_{\D^*,g,{\bf x}_2,\bs{\alpha}_2 }\rangle_{\mc{H}}F_{Q+ip}^{-1}(\nu,\nu')F_{Q+ip}^{-1}(\tilde{\nu},\tilde{\nu}')\dd p.
\end{align*}
The task is therefore reduced to computing the matrix elements of the disk amplitude with two marked points. A key structural feature of these amplitudes is the holomorphic factorization \cite{GKRV20_bootstrap}: 
\begin{equation}\label{matrixcoefampl}
\langle   \mc{A}_{\D,g_0,{\bf x}_1, \bs{\alpha}_1},\Psi_{Q+ip,\nu,\tilde{\nu}}\rangle_{\mc{H}}=w(\Delta_{\alpha_1},\Delta_{\alpha_2},\Delta_{Q+ip},\nu,{\bf x}_1)
\overline{w(\Delta_{\alpha_1},\Delta_{\alpha_2},\Delta_{Q+ip},\tilde{\nu},{\bf x}_1)}
\langle   \mc{A}_{\D,g_0,{\bf x}_1, \bs{\alpha}_1},\Psi_{Q+ip}\rangle_{\mc{H}}
\end{equation}
where the coefficients $w(\cdot)$ are holomorphic in ${\bf x}_1$ and can be obtained iteratively by applying linear differential operators with holomorphic coefficients to the three-point function \eqref{3pointDOZZ}, as will be explained below. A similar factorization holds for $\langle \Psi_{Q+ip,\nu',\tilde{\nu}'}, \mc{A}_{\D^*,g,{\bf x}_2,\bs{\alpha}_2 }\rangle_{\mc{H}}$.

To compute the remaining factor $\langle   \mc{A}_{\D,g_0,{\bf x}_1, \bs{\alpha}_1},\Psi_{Q+ip}\rangle_{\mc{H}}$, we use that the map $\alpha \in \C\mapsto \langle   \mc{A}_{\D,g_0,{\bf x}_1, \bs{\alpha}_1},\Psi_{\alpha}\rangle_{\mc{H}}$ is anti-holomorphic. In the probabilistic regime $\alpha<Q$, one has $\Psi_{\alpha}=C(g)\mathcal{A}_{\D,g,0,\alpha}$, the amplitude of a disk with a marked point at $0$ and weight $\alpha$ (up to a metric-dependent factor $C(g)$, the exact value of which will be skipped for simplicity).  Composing the amplitudes of two such disks  gives a three-point correlation function on the sphere, so that by \cref{3pointDOZZ}:
\begin{align*} 
\langle   \mc{A}_{\D,g_0,{\bf x}_1, \bs{\alpha}_1},\Psi_{\alpha}\rangle_{\mc{H}}=&C(g) \langle V_{\alpha_1}(0)V_{\alpha_2}(z)V_{\alpha}(\infty)\rangle_{\mathbb{S}^2,g}\\
=& C(g)C_{\gamma,\mu}^{\rm DOZZ}(\alpha,\alpha_1,\alpha_2)|z|^{2(\Delta_\alpha-\Delta_{\alpha_1}-\Delta_{\alpha_2})}g(z)^{-4\Delta_{\alpha_2}}
\end{align*}
and this extends anti-holomorphically to $\alpha=Q+ip$. In the same way, we get  
$$\langle \Psi_{Q+ip}, \mc{A}_{\D^*,g,{\bf x}_2,\bs{\alpha}_2 }\rangle_{\mc{H}}=C(g)2^{2\Delta_{\alpha_4}-2\Delta_{\alpha_3}-2\Delta_{\alpha_{Q+ip}}}g(2)^{-\Delta_{\alpha_3}}C_{\gamma,\mu}^{\rm DOZZ}(Q+ip,\alpha_3,\alpha_4) .$$
Combining these computations yields the following:
\begin{theorem}\cite{GKRV20_bootstrap}
The $4$-point correlation function on the sphere can be expressed as 
\begin{equation}
\langle \prod_{i=1}^4 V_{\alpha_i}(x_i)\rangle_{\mathbb{S}^2,g}= C(g) g(2)^{-\Delta_{\alpha_3}}g(z)^{- \Delta_{\alpha_2}}\int_{\R} \overline{C_{\gamma,\mu}^{\rm DOZZ}(Q+ip,\alpha_1,\alpha_2)}C_{\gamma,\mu}^{\rm DOZZ}(Q+ip,\alpha_3,\alpha_4)|\mc{F}_{p,\bs{\alpha}}(z)|^2\dd p
\end{equation}
where $\mc{F}_{p,\bs{\alpha}}(z)$ is the \textbf{Conformal Block}, given by 
 \begin{align}
\mc{F}_{p,\bs{\alpha}}(z)=& 2^{\Delta_{\alpha_4}-\Delta_{\alpha_3}-\Delta_{Q+ip}}z^{\Delta_{Q+ip}-\Delta_{\alpha_1}-\Delta_{\alpha_2}}\\
& \times \sum_{\nu\in \mathcal{T}}F_{Q+ip}^{-1}(\nu,\nu')w(\Delta_{\alpha_1},\Delta_{\alpha_2},\Delta_{Q+ip},\nu,0,z)w(\Delta_{\alpha_3},\Delta_{\alpha_4},\Delta_{Q+ip},\nu',2,\infty) .\nonumber
\end{align}
\end{theorem}

A central tool for establishing holomorphic factorization is the family of Ward identities, which encode the infinitesimal action of conformal symmetries on amplitudes. To keep the exposition simple, we focus on the case of a disk with two marked points.

Let $v$ be a holomorphic vector field near $\D^*$ with flow $f_t$. This induces conformal deformations of the surface, with  $\D_t^*=f_t(\D^*)$, boundary parametrization $\zeta_t:=f_t\circ \zeta$, metric $g_t:=(f_t)_*g$, , and marked points transported as ${\bf x}_t=(f_{-t}(x_3),f_{-t}(x_4))$. By diffeomorphism invariance of amplitudes,
$$
\mc{A}_{\D^*_t,g_t,{\bf x},\bs{\alpha},\zeta_t}= \mc{A}_{\D^*,g,{\bf x}_t,\bs{\alpha},\zeta}.
$$
Moreover, $\D^*_t$ can be viewed as the gluing of a fixed disk $r\D^*\subset \D^*$ (for some $r>1$) with a varying annulus $\A_t:=\D^*_t\setminus r\D^*$. The amplitude therefore decomposes as the composition of those associated with $r\D^*$ and $\A_t$, and the variations of the latter are controlled by \Cref{varyann}. Differentiating at $t=0$ gives the following identity:
 \begin{lemma}[\textbf{Ward identity}]\label{lemma:ward}\cite{PapierBlocs}
If $v=-\sum_{n\in \Z}v_{n}z^{n+1}\partial_z$ is a holomorphic vector field defined on a neighborhood of $\D^*$ in $\mathbb{S}^2$, then  
 \[
  \partial_t \mc{A}_{\D^*,g,{\bf x}_t,\bs{\alpha},\zeta}|_{t=0}=- \mc{A}_{\D^*,g,{\bf x},\bs{\alpha},\zeta} {\bf H}_{v}  +\sum_{j=3}^4\Delta_{\alpha_j}(v'(x_j)+\bar{v}'(x_j)) \mc{A}_{\D^*,g,{\bf x},\bs{\alpha},\zeta}
 .
\]
 \end{lemma}
  
Taking $v=-z^{-n+1}\partial_z$ with $n\geq 1$, one has $2\mathbf{L}_{-n}=\mathbf{H}_v-i\mathbf{H}_{iv}$. Writing $f_t^v$ and $f_t^{iv}$ for the corresponding flows, Lemma \ref{lemma:ward} yields
\begin{align*}
 -2\langle \mathbf{L}_{-n}\Psi_{Q+ip,\nu,\tilde{\nu}'},\mc{A}_{\D^*,g,{\bf x}, \bs{\alpha},\zeta}\rangle_{\mc{H}} =& \partial _t\langle  \Psi_{Q+ip,\nu,\tilde{\nu}'},\mc{A}_{\D^*,g,f_{-t}^v({\bf x}), \bs{\alpha},\zeta}\rangle_{\mc{H}} |_{t=0}  -i\partial _t\langle  \Psi_{Q+ip,\nu,\tilde{\nu}'},\mc{A}_{\D^*,g,f_{-t}^{iv}({\bf x}), \bs{\alpha},\zeta}\rangle_{\mc{H}} |_{t=0}
\\
& -2\sum_{j=3}^4\Delta_{\alpha_j}v'(x_j) \mc{A}_{\D^*,g,{\bf x},\bs{\alpha},\zeta}
\end{align*}  
For $x_j(t):=x_j+tx_j^{-n+1}+o(t)$, $j=3,4$, and with $\nu=\tilde{\nu}=\emptyset$, this becomes
\begin{align}\label{eval_descendant}
\langle  \Psi_{Q+ip,n,\emptyset}, &\mc{A}_{\D^*,g,{\bf x}, \bs{\alpha},\zeta}\rangle_{\mc{H}} =C(g)\sum_{j=3,4}\Big(\frac{(n-1)\Delta_{\alpha_j}}{x_j^n}-\frac{\partial_{x_j}}{x_j^{n-1}}\Big)
 \langle V_{Q+ip}(0)V_{\alpha_3}(x_3)V_{\alpha_4}(x_4)\rangle_{\mathbb{S}^2,g}. 
\end{align}
Inserting the DOZZ formula \eqref{3pointDOZZ} for the three-point function shows that
\[
\langle  \Psi_{Q+ip,n,\emptyset},\mc{A}_{\D^*,g,{\bf x}, \bs{\alpha},\zeta}\rangle_{\mc{H}} =w(\Delta_{\alpha_3},\Delta_{\alpha_4},\Delta_{Q+ip},n,{\bf x}) 
\langle  \Psi_{Q+ip },\mc{A}_{\D^*,g,{\bf x}, \bs{\alpha},\zeta}\rangle_{\mc{H}} 
\]
for an explicit   coefficient $w(\Delta_{\alpha_3},\Delta_{\alpha_4},\Delta_{Q+ip},n,{\bf x}) $ holomorphic in $x_3,x_4$. A similar argument with $\tilde{\bf L}_{-n}$ instead of ${\bf L}_{-n}$, using $+i\partial_t$ in \eqref{eval_descendant} and , involving the conjugate operator
$$\frac{(n-1)\Delta_{\alpha_j}}{\bar{x}_j^n}-\frac{\partial_{\bar{x}_j}}{\bar{x}_j^{n-1}},$$
 gives the conjugate relation for $\Psi_{Q+ip,\emptyset,n}$.
Iterating these identities leads to the holomorphic factorization of \eqref{matrixcoefampl} for $\langle \Psi_{Q+ip,\nu,\tilde{\nu}'}, \mc{A}_{\D^*,g,{\bf x}_2,\bs{\alpha}2 }\rangle_{\mc{H}}$.

The Ward identities extend to amplitudes with arbitrary numbers of boundary components and marked points \cite{PapierBlocs}, and hence imply holomorphic factorization for all pant amplitudes. They are not specific to Liouville theory but a universal feature of CFT, encoding conformal symmetry. The approach presented here is reminiscent of the Vertex Operator Algebra  formalism, although the derivation via \Cref{sec:annulus} relies on a probabilistic representation of annuli specific to Liouville theory. Yet, the original approach of \cite{GKRV21_Segal} used a probabilistic representation of the stress-energy tensor (SET), formally
$$
T(z):=Q\partial^2_{zz}\phi(z)-(\partial_z\phi)^2(z),
$$
and expressed Virasoro generators as contour integrals of   $T(z)$-insertions. Ward identities then follow from Gaussian integration by parts \cite{GKRV20_bootstrap,GKRV21_Segal}. For a pedagogical introduction to the SET in Liouville theory, see \cite{OikarinenK,Oikarinen}. The SET perspective is discussed in greater detail in the review  \cite{GKRreview}.

\begin{figure}
\begin{center}
 \begin{tikzpicture}
 \node[inner sep=0pt] (pant) at (0,0)
{\includegraphics[width=0.3\textwidth]{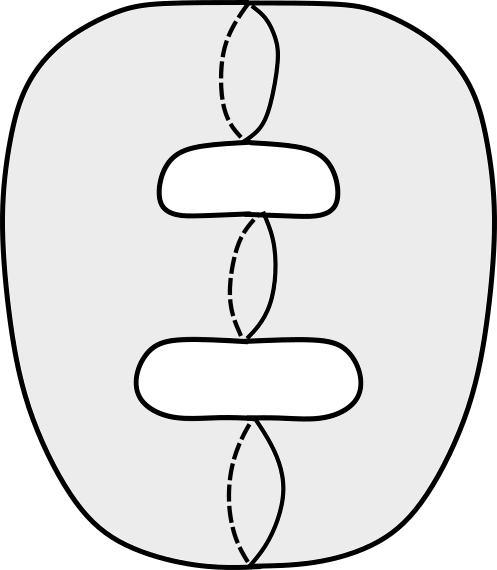}};
 \draw (-2,0) node[right,black]{$\Sigma^1$} ;
  \draw (1.5,0) node[right,black]{$\Sigma^2$} ; 
   \draw (0.2,0) node[right,black]{$\mc{C}_2$} ;
      \draw (0.2,2) node[right,black]{$\mc{C}_3$} ;
        \draw (0.3,-2) node[right,black]{$\mc{C}_1$} ;
\end{tikzpicture}
 \caption{A genus $2$ surface cut into two pairs of pants $\Sigma^1,\Sigma^2$ along $3$ curves 
        $\mc{C}=(\mc{C}_1,\mc{C}_2,\mc{C}_3)$.}\label{Genus2}
        \end{center}
\end{figure}
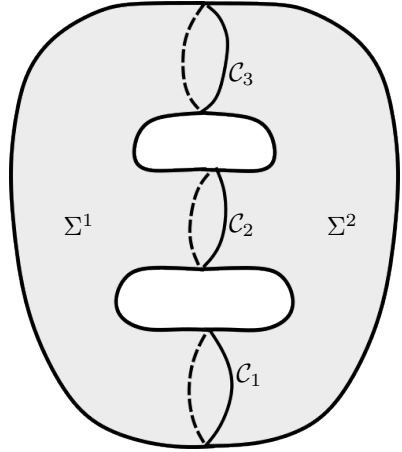

\bigskip 
\paragraph*{Partition function of a genus two surface.} 
We consider a surface $\Sigma$ of genus $2$ that we cut along three curves $\mc{C}=(\mc{C}_1,\mc{C}_2,\mc{C}_3)$, parametrized by $\bs{\zeta}=(\zeta_1,\zeta_2,\zeta_3)$, and we obtain two pairs of pants $\Sigma^1,\Sigma^2$ with parametrized boundary -- See Figure \ref{Genus2}. Assume that all the boundaries are incoming in $\Sigma^1$ and consider a metric $g$ on $\Sigma$ which behaves smoothly with respect to this cutting.

Segal's gluing  leads to
\begin{equation}\label{split_genus2}
Z_{\Sigma,g}=\mc{A}_{\Sigma,g}=\langle \mc{A}_{\Sigma^1,g,\bs{\zeta}},\mc{A}_{\Sigma^2,g,\bs{\zeta}}\rangle_{\mc{H}^{\otimes 3}}
\end{equation}
where we view both  $\mc{A}_{\Sigma^j,g,\bs{\zeta}}$ as elements in $\mc{H}^{\otimes^3}$. For each $\mc{H}\leftrightarrow \mc{H}$ 
pairing above, we use the spectral decomposition  \cref{spectral} and get 
\begin{align}
Z_{\Sigma,g}=& \frac{1}{(2\pi)^3}\int_{\R_+^3}\sum_{\bs{\nu},\bs{\nu}' ,\tilde{\bs{\nu}},\tilde{\bs{\nu}}'\in \mc{T}^3}\langle \mc{A}_{\Sigma^1,g,\bs{\zeta}},\otimes_{j=1}^3\Psi_{Q+ip_j,\nu_j,\tilde{\nu}_j}\rangle_{\mc{H}^{\otimes 3}}\langle\otimes_{j=1}^3\Psi_{Q+ip_j,\nu_j',\tilde{\nu}_j'} ,\mc{A}_{\Sigma^2,g,\bs{\zeta}}\rangle_{\mc{H}^{\otimes 3}}\\
&  \qquad \qquad\quad  \times\prod_{j=1}^3F^{-1}_{Q+ip_j}(\nu_j,\nu_j')F^{-1}_{Q+ip_j}(\tilde{\nu}_j,\tilde{\nu}_j')\dd p_1\dd p_2\dd p_3
\end{align}
where $\bs{\nu}=(\nu_1,\nu_2,\nu_3)$ and similarly $\bs{\nu},\tilde{\bs{\nu}},\tilde{\bs{\nu}}'$. The holomorphic factorization produces here
\begin{equation}\label{descendant_matrix}
\langle \mc{A}_{\Sigma^1,g,\bs{\zeta}},\otimes_{j=1}^3\Psi_{Q+ip_j,\nu_j,\tilde{\nu}_j}\rangle_{\mc{H}^{\otimes 3}}=
C(\Sigma^1,g,\bs{\Delta}_{{\bf Q}+i{\bf p}})w(\Sigma^1,\bs{\Delta}_{{\bf Q}+i{\bf p}},\bs{\nu})\overline{w(\Sigma^1,\bs{\Delta}_{{\bf Q}+i{\bf p}},\tilde{\bs{\nu}})}\,\overline{C_{\gamma,\mu}^{\rm DOZZ}({\bf Q}+i{\bf p})}
\end{equation}
where ${\bf Q}+i{\bf p}:=(Q+ip_1,Q+ip_2,Q+ip_3)$, $\bs{\Delta}_{\bs{Q}+i{\bf p}}:=(\Delta_{Q+ip_1},\Delta_{Q+ip_2},\Delta_{Q+ip_3})$ and $w (\Sigma^1,\bs{\Delta}_{{\bf Q}+i{\bf p}},\bs{\nu})$ are coefficients depending in a holomorphic way on the  complex structure of $\Sigma^1$, and $C(\Sigma^1,g,\bs{\Delta}_{{\bf Q}+i{\bf p}})>0$ is a simple explicit function of its parameters, the exact value of which we skip for simplicity. The same holds for  $\Sigma^2$ 
\[\langle \otimes_{j=1}^3\Psi_{Q+ip_j,\nu'_j,\tilde{\nu}'_j},\mc{A}_{\Sigma^2,g,\bs{\zeta}}\rangle_{\mc{H}^{\otimes 3}}=
C(\Sigma^2,g,\bs{\Delta}_{{\bf Q}+i{\bf p}})w(\Sigma^2,\bs{\Delta}_{{\bf Q}+i{\bf p}},\bs{\nu}')\overline{w(\Sigma^2,\bs{\Delta}_{{\bf Q}+i{\bf p}},\tilde{\bs{\nu}}')}C_{\gamma,\mu}^{\rm DOZZ}({\bf Q}+i{\bf p}).
\]
Combining what we just discussed, we obtain
\begin{theorem}\cite{GKRV21_Segal}
The partition function of the genus $2$ surface $(\Sigma,g)$ cut along the curves $\mc{C}$ parametrized by $\bs{\zeta}$ can be written under the form
\[Z_g= \frac{1}{(2\pi)^3}\int_{\R_+^3}|C_{\gamma,\mu}^{\rm DOZZ}({\bf Q}+i{\bf p})|^2|\mc{F}_{{\bf p}}(\Sigma,\bs{\zeta})|^2\dd p_1\dd p_2\dd p_3 \]
where $\mc{F}_{\bf p}(\Sigma,\bs{\zeta})$ is the \textbf{Conformal Block} of $\Sigma$ defined by 
\[\mc{F}_{\bf p}(\Sigma,\bs{\zeta})=C'(\Sigma^1,\Sigma^2, g,\bs{\Delta}_{\bs{Q}+i{\bf p}} )\sum_{\bs{\nu},\bs{\nu}'\in \mc{T}^3}w(\Sigma^1,\bs{\Delta}_{{\bf Q}+i{\bf p}},\bs{\nu})
w(\Sigma^2,\bs{\Delta}_{{\bf Q}+i{\bf p}},\bs{\nu}')\prod_{j=1}^3F^{-1}_{Q+ip_j}(\nu_j,\nu_j')\]
and $C'(\Sigma^1,\Sigma^2, g,\bs{\Delta}_{\bs{Q}+i{\bf p}} )$ is an ``holomorphic square root'' of the product $C(\Sigma^1,g,\bs{\Delta}_{{\bf Q}+i{\bf p}})C(\Sigma^2,g,\bs{\Delta}_{{\bf Q}+i{\bf p}})$.
\end{theorem}

The convergence of the conformal block for almost every ${\bf p}$ follows from the argument in \cite{GKRV21_Segal}, and had not been established before. The constant  $C'(\Sigma^1,\Sigma^2, g,\bs{\Delta}_{\bs{Q}+i{\bf p}} )$ is holomorphic in the complex structures of the pairs of pants $\Sigma^1$ and $\Sigma^2$, as well as in the conformal weights. It satisfies
$$|C'(\Sigma^1,\Sigma^2, g,\bs{\Delta}_{\bs{Q}+i{\bf p}} )|^2=C(\Sigma^1,g,\bs{\Delta}_{{\bf Q}+i{\bf p}})C(\Sigma^2,g,\bs{\Delta}_{{\bf Q}+i{\bf p}}),$$
so it naturally appears as a holomorphic square root of the product on the right-hand side. Selecting such a holomorphic square root (with respect to the complex structure) is a delicate matter, and leads to viewing conformal blocks as holomorphic sections of a line bundle over the Teichm\"uller space of Riemann surfaces with marked points.
Although the blocks depend on the choice of cutting curves $\bs{\zeta}$, this dependence occurs only through a Schwarzian cocycle, which in turn induces a projective unitary action of the Mapping Class Group on the space of conformal blocks. For clarity, we omit the precise definition of $C'$ here and refer the reader to \cite{PapierBlocs}.

\section{Further developments, related results and open problems.}

Beyond the probabilistic construction and bootstrap of Liouville CFT, many directions have emerged at the crossroads of probability, geometry, and physics. We briefly review some of these developments,   and open challenges.

\medskip

\paragraph*{Exact results for GMC theory and boundary Liouville CFT.}
Several exact formulas for Gaussian multiplicative chaos (GMC) measures in one dimension were first conjectured in the physics literature, especially in the study of disordered systems.  A central example is the explicit formula of Fyodorov--Bouchaud  \cite{Fyodorov_2008} for the moments of the total mass of GMC on the circle (see also  Fyodorov-Le Doussal-Rosso \cite{Fyodorov_2009} and Ostrovsky \cite{Ostrovsky_2009, Ostrovsky_2018} for extensions to other geometries). Remarkably, this formula turns out to be a special case of the boundary Liouville structure constants. These structure constants were fully determined in a series of works by Ang-R\'emy-Sun-Zhu \cite{AngRemySun21_FZZ,ang2023derivation}. A particularly transparent instance is R\'emy's CFT-based derivation of the Fyodorov--Bouchaud formula \cite{Remy20}, later extended with Zhu to the case of an interval \cite{Remy_2020}. 

Wu initiated in  \cite{Wu} the study of the conformal bootstrap for surfaces with a boundary in the case of the annulus. Building on this work, the law of the random modulus    for Random Planar Maps with the topology of an annulus was obtained by Ang-Remy-Sun in \cite{Xin}. A broader framework aiming at developing the conformal bootstrap for boundary Liouville CFT was proposed in \cite{Liouvilleboundary}, establishing the validity of Segal's axioms in this context.

\medskip

\paragraph*{Connections to other statistical physics models.} The structure constants in Liouville CFT, i.e. the DOZZ formula, is expected to encode many other statistical physics models in some way. This has been highlighted via the powerful interplay between Liouville CFT and  Schramm Loewner Evolution \cite{Schramm}, or Conformal Loop Ensembles \cite{Sheffield-Werner}, which can be coupled via the Mating of Tree formalism \cite{MatingOfTrees,nina}, a framework developed by Duplantier-Miller-Sheffield  to study the coupling between SLE or CLE  and the GFF in terms of more classical stochastic processes. This has led to   unexpected results for various statistical physics models, like the computation of the three point correlation functions of Conformal Loop Ensembles  \cite{AngSun21_CLE}, shown to be given by the imaginary DOZZ formula,  or the backbone exponent in critical percolation  \cite{nolin}.

 \medskip
 
\paragraph*{Conformal blocks.}

In \cite{PapierBlocs}, it is proven that the conformal blocks are holomorphic sections of a holomorphic line bundle over Teichm\"uller space of Riemann surfaces with marked points, laying the foundations  for a representation of the mapping class group in the space of conformal blocks. A quantization of the Teichm\"uller space using hyperbolic geometry was constructed by Chekhov, Fock \cite{CF} and Kashaev \cite{kashaev}. The Liouville conformal blocks conjecturally provide another such quantization.  Verlinde \cite{VERLINDE1990652} conjectured their equivalence, and Teschner \cite{Teschner:2002vx,TeschnerTeich} provided strong arguments for this. The   probabilistic construction of Liouville conformal blocks \cite{GKRV21_Segal,blocguillaume} paves the way towards a proof of this correspondence. These quantizations are related through non-compact quantum groups representation theory, whose connection for Liouville was rigorously established in \cite{blocguillaume}.

Some specific conformal blocks of LCFT have been shown to admit probabilistic expressions involving statistical moments of GMC \cite{blocguillaume}. The question whether this representation holds in a more systematic way is under investigation.

The AGT correspondence \cite{Alday_2010} between $4d$ supersymmetric Yang-Mills theory and the bootstrap construction of LCFT conjectures  that Liouville conformal blocks coincide  with special cases of Nekrasov's partition function \cite{Nekrasov2003}. In particular this leads to  an explicit formula for the coefficients in the series expansion of the conformal blocks in the moduli parameter. However, even admitting this conjecture, it remains difficult to control   the radius of convergence:  
see for instance  \cite{Felder2018}. 
The AGT conjecture has  been proved  as an identity between formal power series in the case of the torus in  \cite{Negut2016} following the works  \cite{MaulikOkounkov2019, SchiffmannVasserot2013},  but this does not address the issue of convergence.  See also \cite{FateevLitvinov, AlbaFateevLitvinovTarnopolsky2011}  
 for arguments in the physics literature which support the AGT conjecture on the torus or the Riemann sphere.
 
 \medskip 
 
\paragraph*{JT-gravity}
 {\it 2d gravity.} LCFT describes the fluctuating metric in 2d gravity. A related model, Jackiv-Teitelboim or Dilaton gravity \cite{JACKIW1985343,TEITELBOIM198341} has been argued in physics to describe large scale (low energy) behaviour of the SYK model of disordered (fermionic) spins with interesting connections to  random matrix theory.  JT gravity also is argued to compute volumes of the Weil-Petersson measure on the moduli spaces of surfaces with boundaries. JT gravity itself is supposed to be a semiclassical limit of Liouville Quantum Gravity \cite{mertens} and, on surfaces with boundary, it reduces to Schwarzian quantum mechanics on the boundary, which was recently given a probabilistic construction \cite{bauerschmidtST}. There are lots of issues in this circle of ideas to understand from the probabilistic angle, the definition of JT, its relation to LQG and the Schwarzian theory.

\medskip

\paragraph*{Path integrals for other CFTs.}

Liouville CFT fits as a particular case of a more general CFT called Toda CFT, where the field takes values in the Cartan algebra of a semi-simple Lie algebra. This model is also related to the AGT correspondence. The probabilistic construction of the path integral was done recently by Cercl\'e-Rhodes-Vargas \cite{Cercle-Rhodes-Vargas} and by \cite{CH1,CH2,CH3} for the boundary Toda CFT. Cercl\'e  \cite{cercle} has been able to prove the formula of 
Fateev-Litvinov for the $3$-point function for certain range of parameters. The general expression of the $3$-point function is not known.  In physics, it is argued that the symmetry algebra of the Toda CFT is the W-algebra, which is strictly bigger than the Virasoro algebra. Understanding the conformal bootstrap in this context is a challenging open problem.

Recently, the probabilistic construction of a non linear sigma model with a further topological term, where the fields take values in the $3$D
hyperbolic space $H^3 = {\rm SL}(2, C)\setminus {\rm SU}(2)$, was carried out in \cite{GKRH3}. This model belongs to the large family of Wess-Zumino-Witten (WZW)
models, which are among the most studied CFTs in physics. They have a formal path integral representation in terms of fields taking values
in a semisimple Lie group G, or a coset space. Many of their properties have been obtained algebraically from
their postulated affine Lie algebra symmetry, especially in the case of compact G. Yet, their rigorous probabilistic
construction and analysis starting from the path integral has remained a challenge and \cite{GKRH3} provides the first such
construction. Implementing the conformal bootstrap for WZW models is a success of the VOA approach in the case of compact groups G but the case of noncompact groups, or cosets like $H^3$,  is an open problem.

 The Liouville model is believed to make sense for imaginary values of the parameter $\gamma$, in which case it should describe the scaling limit of statistical physics models.  The quest for a probabilistic construction of imaginary Liouville theory  was initiated in \cite{CILT} for a compactified version of this theory. This  work shows a rich structure of (non-unitary) Logarithmic CFT, therefore providing a testbed for the mathematical study of this concept.   The grand challenge is to understand the probabilistic approach of the (non compact) imaginary Liouville theory.

\medskip
\paragraph*{Conclusion.}
The case of Liouville CFT shows that a rigorous path integral construction can be turned into a complete bootstrap program. The roadmap is clear: start from probabilistic expressions of correlation functions in a suitable parameter regime;  apply Segal's axioms to decompose them into amplitudes in a Hilbert space framework (via conditioning on circles);  analyze the propagator, which corresponds to the annulus amplitude, as a spectral problem (scattering in the Liouville case); identify the eigenstates as Virasoro descendants of certain primaries; and finally, express the pair-of-pants amplitudes in this basis. Inserting degenerate fields then determines the three-point coefficients on primaries, while Ward identities extend this to descendants, yielding the conformal blocks. Altogether, these ingredients combine into the full bootstrap formulas.

For more general CFTs, following this roadmap remains a formidable challenge. One must first identify suitable probabilistic representations (as in the examples discussed above), then tackle the spectral analysis, which is particularly delicate in non-unitary or continuous-spectrum theories (for instance Toda or $H^3$-WZW). The presence of extended symmetry algebras (W-algebras, affine Lie algebras, supersymmetry, etc) adds further layers of complexity, and the degenerate fields may not determine the full three point functions (like in Toda).
It is also natural to ask whether probabilistic, VOA, and topological recursion approaches can be combined to obtain a more comprehensive picture. At a broader level, it seems natural to expect that the mapping class group representations produced by conformal blocks should admit geometric incarnations, both in terms of quantization of moduli spaces and in relation to 3D topological quantum field theory, as exemplified by the Liouville-Teichm\"uller correspondence and by WZW models with compact groups \cite{Teschner:2002vx,TeschnerTeich,VERLINDE1990652}. The real question is how far such correspondences extend, and in what precise form they manifest in more general, non-rational or non-compact CFTs.

 Finally, the probabilistic couplings between Liouville theory and random curves such as SLE and CLE have already proved remarkably fruitful; designing their generalizations to more complex models is another promising direction. Thus, while Liouville CFT provides the first complete example, it also suggests a conceptual blueprint for other theories. Whether this program can be carried through for models such as Toda, WZW, logarithmic CFTs, or coupled systems like Liouville-CLE/SLE remains an open and exciting challenge for the years to come.

\section*{Acknowledgments.}
 It is a pleasure to thank all our collaborators for the many insights and ideas we have shared over the years, in particular Guillaume Baverez, Colin Guillarmou and Antti Kupiainen, whose contributions and perspectives have been invaluable to this project. We are also very grateful to  Colin Guillarmou,  Yulai Huang, Nathan Hughenin, Baojun Wu for carefully reading preliminary versions of this manuscript and for many helpful comments and suggestions that have significantly improved the presentation.

\bibliographystyle{siamplain}
\bibliography{example_references}
\end{document}